\documentclass[sigconf,nonacm]{acmart}

\AtBeginDocument{%
  }

\usepackage{multirow}
\usepackage{enumitem}
\usepackage{caption}
\usepackage{array}
\usepackage{graphicx}
\usepackage{subfig}

\usepackage{graphicx} 
\usepackage{caption}

\newcommand{\etal}{{\em et al.}}
\long\def\omitit#1{}

\begin{document}

\title{Fanfiction in the Age of AI: Community Perspectives on Creativity, Authenticity and Adoption}

\author{Roi Alfassi}
\affiliation{%
  \institution{University of Haifa}
  \city{Haifa}
  \country{Israel}}
\email{roialfassi@gmail.com}

\author{Angelora Cooper}
\author{Zoe Mitchell}
\author{Mary Calabro}
\author{Orit Shaer}
\affiliation{%
  \institution{Wellesley College}
  \city{Wellesley}
  \state{MA}
  \country{USA}
}
\email{acooper5@wellesley.edu}
\email{zm2@wellesley.edu}
\email{mc111@wellesley.edu}
\email{oshaer@wellesley.edu}

\author{Osnat Mokryn*}
\orcid{0000-0002-1241-9015}
\affiliation{%
  \institution{University of Haifa}
  \city{Haifa}
  \country{Israel}}
\email{omokryn@is.haifa.ac.il}

\thanks{\textbf{Accepted for publication in the \textit{International Journal of Human-Computer Interaction}, June 2025}}
\thanks{* Corresponding author. Email: omokryn@is.haifa.ac.il}

\renewcommand{\shortauthors}{Alfassi et al.}

\begin{abstract}
   The integration of Generative AI (GenAI) into creative communities, like fanfiction, is reshaping how stories are created, shared, and valued. This study investigates the perceptions of 157 active fanfiction members, both readers and writers, regarding AI-generated content in fanfiction. Our research explores the impact of GenAI on community dynamics, examining how AI affects the participatory and collaborative nature of these spaces. The findings reveal responses ranging from cautious acceptance of AI's potential for creative enhancement to concerns about authenticity, ethical issues, and the erosion of human-centered values. Participants emphasized the importance of transparency and expressed worries about losing social connections. Our study highlights the need for thoughtful AI integration in creative platforms using design interventions that enable ethical practices, promote transparency, increase engagement and connection, and preserve the community's core values. 
   
\end{abstract}

\omitit{

}
\begin{teaserfigure}
\centering
  \includegraphics[width=.95\textwidth]{ 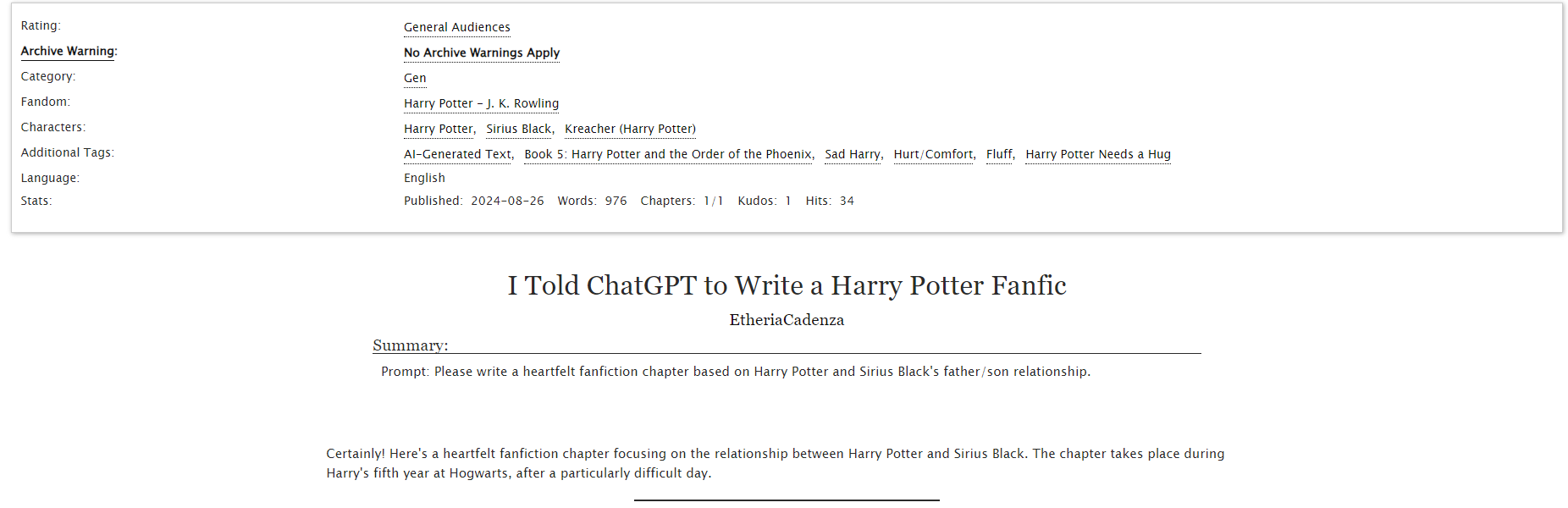}
  \caption{A fanfiction contributor shares a fanfiction story created by ChatGPT in August 2024~\cite{beepboopAI}. }
  \Description{A fanfiction contributor shares a fanfiction story created by ChatGPT in August 2024}
  \label{fig:teaser}
\end{teaserfigure}
\maketitle

\section{Introduction}
Recent advancements in Generative Artificial Intelligence (GenAI) have introduced a technological disruption in creative practices that is not yet fully understood. Human-GenAI co-creative processes have been shown to enhance human creativity~\cite{doshi2024generative,ShaerBrainwriting24}, with several studies highlighting a sense of human empowerment, where individuals described feeling like they have “a second mind”~\cite{wan2024felt}. Educators have also noted the potential for GenAI to foster intellectual emancipation, which involves liberating human thought by challenging norms within a knowledge domain~\cite{tsao2024beyond}. However, GenAI tools have also raised concerns about creation and authorship in creative fields~\cite{epstein2023art,hbr2023bgenerative}. These concerns are amplified by the language used to describe GenAI tools and the language these tools use when interacting with humans, often designed to convey a sense of agency and human-like intent~\cite{epstein2020gets}. 
The human-GenAI co-creation is believed to be influenced nowadays by ``machine culture'', as GenAI tools select which narratives they transmit~\cite{brinkmann2023machine}, and creativity, in writing and storytelling, is reconceptualized as distributed across human and non-human agents~\cite{tsao2024beyond}.
Similarly, concerns were raised for the co-creation writing processes themselves, as reliance on AI may homogenize content~\cite{doshi2024generative}, amplify bias~\cite {Bender2021}, and potentially reduce the originality and depth of human-driven storytelling~\cite{hbr2023bgenerative}. Additionally, the pervasive use of AI risks eroding writers' creative autonomy and storytelling skills and might lead to an over-dependence on AI-generated outputs~\cite{roemmele2018creative}.  

These concerns require investigating the impact of GenAI not only on creative practices but also on communities.  Originally designed for human creativity and collaboration, online spaces are seeing an increased presence of overt and covert intelligent agents that are altering the communities' culture and dynamics~\cite{tsvetkova2024human}. As creative communities evolve and change to include machines, the transition might present challenges to creative communities and their values. This is especially important for communities deeply engaged in a collaborative and iterative creative process that places the human at the center, such as a storytelling one. 

The fanfiction community is a prime example of storytelling and creative expression \cite{hellekson2009fannish,Yin2017}. Fanfiction, or fanfic, refers to stories that are written with the intent to engage with popular stories or media, and are considered as drawing on the tradition of oral storytelling~\cite{jenkins2012textual}.   Fanfiction has its roots in various fandoms that have existed as long as stories have been told, but has evolved, with the introduction of the Web, into vibrant, global networks of non-profit writers and readers~\cite{Yin2017}.  An online fanfiction platform encompasses millions of writers and readers globally. Platforms like Archive of Our Own (AO3) alone host over 14 million works across 68,000+ fandoms \cite{ao3}, demonstrating its expansive reach. Fanfiction culture and platforms also have a significant impact on the book market \cite{Soller_2019, flegel2016}. 
Fanfiction communities are characterized by their strong emphasis on originality, community-driven feedback, and the reimagining of existing narratives~\cite{hellekson2009fannish, Yin2017}. Hence, they provide a unique context for understanding the evolution process of an online creative community into an integrated human-AI one, as well as the forces of resistance and adoption within such a community. Studying fanfiction members' perceptions of this evolution, how readers and writers perceive co-creation with GenAI, intelligent agents' created content, and how they utilize GenAI tools could offer insights into a creative community that is undergoing a transition process. This could also shed light on the processes that such a community undergoes during a phase that might involve augmentation, where GenAI assists human creators, and automation, where the content is created by intelligent agents (e.g., Figure \ref{fig:teaser})~\cite{ali2023using,hbr2023bgenerative}. 

Additionally, these communities often serve as early adopters and critical evaluators of new technologies. Their attitude and experiences with GenAI can inform the design of AI tools and online platforms that are better aligned with the values of creativity, collaboration, appreciation for human labor, and originality that are central to communities of making and arts ~\cite{Poddar2023,  winter2021communities}, and promote our understanding of how AI can be harnessed to support or potentially disrupt not only individuals but also collective creative processes. 

In this paper, we present findings from a study involving 157 active members of fanfiction communities, who are fanfic writers or readers. We explore participants' attitudes and perceptions toward GenAI-created content, their views on AI's role in the community, and in shaping the future of storytelling within fanfiction. We explore writers' knowledge of, and engagement with, AI-assisted writing tools, and their motivations for incorporating (or rejecting) AI in their creative processes. 

We study the complexities of the integration process of GenAI into fanfiction spaces, where storytelling is not only a creative outlet but also a means of building social connections, developing skills, and sharing personal interpretations of fictional worlds and characters. 

By exploring the nuanced attitudes, behaviors, and perceptions of fanfiction participants towards AI, we aim to contribute to the ongoing discourse on the ethical, cultural, and practical dimensions of integrating AI into creative practices, ensuring that the technology evolves in a way that enhances rather than diminishes human creativity.

Our contributions in the paper are as follows. 
\begin{enumerate}
    \item Community's changing dynamics: We explore how the presence of AI-generated content influences the perception of interactions within fanfiction communities. While some participants see potential benefits, such as revitalizing dormant fandoms, many expressed concerns about AI's impact on community values, including creativity, authenticity, and the social connections that define these spaces. Our findings reveal a spectrum of reactions, from neutrality or mild acceptance to active avoidance and disdain, suggesting that GenAI's integration may alter community dynamics and engagement patterns. 
    \item Participatory culture: We study how the integration of AI tools into the writing process at various levels and the possibility of automation are perceived. Our study highlights current use by early adopters alongside concerns about the potential erosion of fanfiction’s participatory culture. These concerns include the risk of weakening the community bonds integral to the fanfiction experience, prompting a need for thoughtful integration of AI that respects the core values of these creative spaces.
    \item Implications for the design of online platforms: The study highlights the need for design interventions that prioritize transparency, ethical integration of AI, and the cultivation of community engagement, while upholding core values such as appreciation for human creativity, labor, and collaboration. We propose design interventions for fanfiction platforms, potentially applicable to other creative communities.
\end{enumerate}

\section{Related Work}
The question of how GenAI is transforming creative communities, and the involved dynamics, is of much interest~\cite{birhane2022power,lamerichs2023generative}. Of particular interest is the fanfiction online community, considered a frontier for gaining insights into the complex interactions and practices among individual users, communities, and intelligent machines~\cite{li2024fandom}.
\subsection{Fanfiction communities as a participatory space}
Fanfiction communities have been studied extensively as spaces that foster collaborative storytelling and creative expression \cite{thomas2011fanfiction}. These participatory~\cite{jenkins2016participatory} communities allow their members to engage deeply with existing narratives, often reimagining, interpreting, interrupting, and expanding upon them in ways that create new themes and character developments~\cite{jamison2013fic}. Thomas notes that the origin of fan fiction can be linked back to ``oral and mythic traditions'' ~\cite{thomas2011fanfiction}. Fanfiction is inherently participatory, with the pleasure of reading fanfiction often seen as stemming from engaging with the narrative. Similarly, fanfiction writers actively reinterpret and expand existing stories in a participatory manner~\cite{barnes2015fanfiction}.

The rise of digital platforms such as FanFiction.net and Archive of Our Own (AO3) has increased the visibility and accessibility of fanfiction communities~\cite{jenkins2016participatory} and their participatory nature, making it easier to both consume (read) and contribute (through writing, editing, commenting) content~\cite{jenkins2016participatory}. 
These platforms have established new norms for the production and consumption of fan works, promoting interaction through reviews and feedback, which are integral to the community’s learning and growth~\cite{tosenberger2008homosexuality,booth2015playing,Frens2018ReviewsMH, aragon2019writers}. Additionally, these communities have often served as early adopters of emerging technologies, integrating tools that enhance storytelling and participation
~\cite{booth2015playing}. This integration and adaptability of new technologies, makes fanfiction communities an intriguing context for examining the impact of GenAI on participatory cultures, and study human–community–machine interactions (HCMI)~\cite{li2024fandom}.

Li and Pang~\cite{li2024fandom} further argue that the use of AI is not entirely new to fandom. They discuss the use of AI to generate new content, for example, the Harry Potter Botfic~\cite{lamerichs2017aidriven} and AI-generated translations~\cite{ABURAYYASH2024100162}, in the context of potential depreciation of the value of fans' labour within the fandom gift economy~\cite{li2024fandom}. Emerging use of AI in fanfiction communities is not limited to content production but could also influence the process and reception of content. For example, a recent study demonstrated the use of machine learning techniques to predict the popularity of fanfiction based on key elements, such as story length and themes~\cite{nguyen2024big}. 

This work highlights the importance of understanding the nuanced role that AI plays in participatory culture and fanfiction communities, affecting not only the content produced but also their creative and social dynamics. Our work aims to contribute to the broader body of research on AI's role in fanfiction communities, as well as the broader body of work on participatory creative communities. We aim to expand the understanding of GenAI's impact on the creative practices of both individuals and communities, focusing on the transition phase of a participatory community into what might evolve into a human-machine creative community.


\subsection{Human-AI co-creation in creative work}
``Creativity is often held up as a uniquely human quality.. a human masterpiece'', observe De Cremer \etal~\cite{hbr2023bgenerative}. However, today we find large language models, trained on extensive human-created datasets, emerging as powerful players in content generation, creativity, and ideation processes~\cite{hbr2023generative,ShaerBrainwriting24,koivisto2023best}. The use of GenAI has been shown to enhance human creativity~\cite{doshi2024generative}, and AI co-created storytelling is used in diverse areas such as engaging children with literature and learning, medical training, and career development~\cite {sun2024exploring,zhang2024mathemyths,fan2024storyprompt,mittenentzwei2024ai,donald2024supporting}.

The use of AI tools and large language models (LLMs) in creative writing and storytelling has been widely studied, exploring both the opportunities and challenges of these technologies in enhancing human creativity~\cite{davis2013human}. GenAI has been shown to assist in various creative tasks, such as idea generation and development~\cite{prewriting2024,ShaerBrainwriting24, girotra2023ideas}, and writers often see GenAI as a co-creator, using its suggestions to introduce new ideas and narrative twists~\cite{yang2022ai, redhead}. AI tools help in narrative development~\cite{wordcraft2022,sittenfeld2024experiment, Tian2024AreLL, plan-and-write}, and tools for character and persona development, like CharacterMeet~\cite{Qin2024CharacterMeetSC} and NarrativePlay~\cite{Zhao2024NarrativePlayAA} have demonstrated that AI can facilitate dynamic character creation and immersive storytelling through interactive, multimodal experiences.


However, despite these advancements in supporting writers, there are significant challenges and concerns regarding using GenAI in creative writing. It was shown that GenAI support for writers goes beyond improving efficiency to also influencing the direction and diversity of content~\cite{ShaerBrainwriting24, Poddar2023, wordcraft2022, opinionated2023}. GenAI writings reinforce existing biases~\cite {Bender2021}, and lead to increasingly homogenized content generation~\cite{doshi2024generative, Bender2021}.  Expert writers found that AI writing assistants failed to maintain coherent narrative and voice~\cite{Bender2021, ippolito2022, chen2023chatgpt, redhead} and lacked emotional depth~\cite{sittenfeld2024experiment}. Concerns have also been raised about the datasets used to train LLMs, which often include copyrighted material without the consent of the original authors, leading to potential intellectual property challenges~\cite{sittenfeld2024experiment, frenkel2023data}.

To address these challenges, researchers emphasize the need for user-centered design of more effective AI tools for creative writing~\cite{plan-and-write, design_space}. Smith \etal~\cite{design_space} proposed a comprehensive design space framework that categorizes key aspects of intelligent writing assistants, including \textit{task}, \textit{user}, \textit{technology}, \textit{interaction}, and \textit{ecosystem}. This framework offers a structured approach to developing more effective and ethical writing tools. In this aspect, we contribute here to the development of a more nuanced understanding of the aspects of task, user, and ecosystem when considering the use and the development of intelligent writing assistants for creative participatory communities.

Most importantly, we explore here the community's perspective on the future of human-GenAI collaboration. De Cremer \etal~\cite{hbr2023bgenerative} see three alternatives for ways that AI could disrupt creative work, and within it creative writing. One involves the AI supporting human creativity, reducing time and effort. The second involves the danger of cheaply made AI flooding the creative landscape, undercutting authentic human voices, while allowing more personalized content. In this scenario, humans would be mainly curators of AI-generated content. In the third scenario, the uniqueness of human voices is recognized~\cite{shumailov2024ai}.  Here, we ask both readers and writers, some of whom are already experienced with GenAI tools, about their views on the future of content generation with AI. 

\subsection{Ethical concerns in Data Use and AI-Generated content}
Discussions around the ethical use of data in generative AI systems have become increasingly urgent, especially in creative contexts. The boundaries between participation, authorship, and ownership are blurred when models are trained on large corpora that include personal narratives, artistic work, or culturally situated expressions~\cite{gero2025creative}. This has the potential to be particularly sensitive in fanfiction communities, where content is often produced not for profit, but as a form of social connection or identity expression~\cite{jenkins2016participatory,Jenkins2019ArtHN}.

Standard notions of consent don't always apply well in these cases. When data is collected in bulk — often through APIs or web scraping — individuals rarely know that their contributions might become training material~\cite{fiesler2018participant}. Some researchers suggest that consent should be reconceptualized as a dynamic and revocable process, mainly when applied to creative labor that wasn’t originally intended for algorithmic consumption~\cite{hand2018aspects,proferes2021studying}.

Gero~\etal~\cite{gero2025creative} found that writers who have shared their work online often describe a sense of disempowerment when they learn that their stories or language styles might have been absorbed into commercial models. For them, the issue isn’t just about data—it’s about recontextualizing their identity or creative effort without recognition. These responses suggest that traditional definitions of harm may be too narrow to capture the affective and cultural stakes involved in datafication of creative expression~\cite{gero2025creative, proferes2021studying}.

Ethical concerns also intersect with community dynamics. In platforms like fanfiction forums or Reddit, writing is often part of reciprocal exchange—commenting, mentoring, remixing. As generative tools become more common, researchers have asked whether these social structures might weaken~\cite{proferes2021studying}. To this end, our study will explore both readers’ and writers’ perspectives on what constitutes appropriate use of AI in the fanfiction writing process.

Solutions proposed across disciplines range from technical fixes to systemic reforms. Some suggest privacy-enhancing technologies that allow users to control which aspects of their data are shared. Others advocate for platform-level governance, including opt-in/opt-out tools and greater transparency around data use. A few scholars even argue for applying IRB-style oversight models to AI training practices—a move that would reframe dataset curation as a form of human subjects research~\cite{hand2018aspects,ray2023chatgpt}.

Recent discussions in AI ethics increasingly emphasize that creative output should not be treated as neutral raw material, but rather as the outcome of human labor embedded in cultural, social, and emotional contexts~\cite{jo2020lessons}. When such work is reappropriated as training data for AI, it often undergoes decontextualization, resulting in the erosion of agency and the dissociation of the content from its original intent~\cite{gero2025creative}. This process has been described as a form of data colonialism, where creative contributions are extracted and commodified without the creators’ consent or participation in the systems that profit from them~\cite{couldry2020costs}.

These concerns resonate with longstanding debates in participatory media theory, where authorship is frequently collaborative, and value is understood less in financial terms and more in relation to community standing and shared meaning-making~\cite{jenkins2006convergence}. The rise of generative AI challenges these frameworks, especially when authorship and attribution become blurred or obscured. In response, researchers and technologists have proposed governance mechanisms such as consent-aware datasets and opt-in/opt-out registries to better align data practices with ethical expectations~\cite{balan2023decorait,birhane2022power}. Still, despite these technical proposals, many creators report feeling sidelined from the decisions that shape how their contributions are used, raising questions about representation, legitimacy, and the governance of creative labor in AI development~\cite{kyi2025governance,lovato2024foregrounding}.


\subsection{Theoretical foundations: Participatory culture, creative labor, and AI ethics}
\label{sec:theory}
This study is situated at the intersection of three theoretical frameworks: participatory culture, creative labor, and the ethics of artificial intelligence. Together, these perspectives offer a conceptual foundation for examining how fanfiction communities respond to the introduction of generative AI technologies.

Participatory culture is featured by collaborative authorship, shared norms, and collective knowledge production within decentralized creative ecosystems~\cite{jenkins2006convergence,Jenkins2019ArtHN}. Fanfiction communities exemplify these dynamics by emphasizing peer feedback, community recognition, and accessible modes of creative expression. Fanfiction is seen by its members as a socially embedded and emotionally supportive practice~\cite{kelley2021loving}, raising questions of how AI might shape not only individual authorship, but also community relationships and identity formation~\cite{subin2024fanfiction}. 
 Unlike commercial authorship, fanfiction is typically produced without financial compensation, instead motivated by passion, recognition, and community involvement. The unconsented use of such works in AI training~\cite{gero2025creative} sets in motion a form of appropriation that some scholars describe as data colonialism~\cite{couldry2019data}, where the outputs of unpaid, informal labor are extracted and repurposed within commercial and technological systems. This process raises critical concerns regarding attribution, consent, and the devaluation of labor in digital environments~\cite{gero2025creative}.

Our study also engages with debates in AI ethics, particularly within HCI and critical data studies. Scholars in this field have increasingly challenged the notion of data neutrality, pointing to the social, political, and cultural biases embedded within algorithmic systems~\cite{craig2022ai,weerts2024neutrality}. In the context of fanfiction, ethical concerns around provenance, authorship, and consent become especially salient. Generative models that produce content in the style of or derived from fan-authored work may disrupt community norms, destabilize notions of authorship, and provoke tensions around legitimacy and voice~\cite{gero2025creative}. These developments call for a closer examination of how AI is perceived and negotiated within participatory creative cultures.

\section{Study}
\subsection{Goals and research questions}
The aim of our study is to explore the experiences, attitudes, and perceptions of AI and Large Language Models (LLMs) within the fanfiction community and identify processes and dynamics of adoption and resistance. We seek to understand how fanfiction readers and writers perceive GenAI-integration into the writing process, the differences they perceive between human and AI-generated stories and content, how writers interact with AI-powered tools, and the overall impact of these technologies on creative practices within the community. Specifically, we aim to address the following research questions:
\subsubsection{RQ1: Understanding the characteristics of the Fanfiction Community Members Survey Participants} 
What are the characteristics, engagement levels, types, and motivations of fanfiction readers and writers participating in this study? This question aims to better contextualize the perspectives and experiences shared by study participants, and help understand how the ways and tools with which they interact and engage with the content. 
\subsubsection{RQ2: How Do Fanfiction Community Members Perceive Human Versus AI-Generated or AI-Augmented Content, and How Do These Perceptions Influence Community Dynamics?}
This question explores the reception of human-created versus AI-augmented or AI-generated content within the fanfiction community, focusing on participants' perspectives on the creative space and their views on current dynamics.  We examine perceptions of the importance of storytelling factors such as quality, emotional depth, narrative coherence, and authenticity, as well as how these elements influence the overall perception and acceptance of AI-generated or collaborative fanfiction in the space and the dynamics of the community.
\subsubsection{RQ3: How are AI and LLMs perceived within the fanfiction community?}
How are AI tools and Large Language Models perceived by the fanfiction community, particularly in the context of creative writing and storytelling? How these tools are viewed in terms of their capabilities, potential benefits, and perceived threats to creativity,  originality, and changing dynamics.

\subsection{Method}
\subsubsection{Questionnaire}
To address these questions, we decided to employ a structured online questionnaire. This method was chosen for several reasons. First, participant anonymity is a core characteristic of interactions within fanfiction communities, allowing individuals to freely share their perspectives without fear of judgment or exposure \cite{BaconSmith1992EnterprisingWomen, Cheng22}. An anonymous survey aligns with these norms, ensuring participants feel comfortable discussing their views. Second, we sought to capture a diverse range of perspectives by reaching a relatively large number of participants across various platforms with varying levels of engagement and roles within the community. Lastly, the fanfiction community has a well-documented affinity for written engagement and communication, whether through storytelling, reviews, or discussions \cite{jamison2013fic, thomas2011fanfiction, Cheng22}. A written questionnaire leverages this affinity and is aligned with the community's established norms and practices.

Similarly to the approach described in a study on gaming perceptions \cite{ShaerCHI17}, we designed our questionnaire to explore demographics, habits, attitudes, and opinions. The survey was designed to address both fanfiction readers and writers. The questions were presented in multiple formats, including multiple-choice, Likert scale, and open-ended formats, allowing for both structured and narrative responses. To ensure consistency and participant attention, we randomized the order of certain questions exploring attitudes and perceptions and included strategically repeated questions to check for response reliability.

We collected demographic information such as age, gender, education, and country. We then asked participants to identify as fanfiction readers, writers, or both. The majority of the questionnaire addressed readers, and 26 questions addressed writers exclusively. Table~\ref{questP1} includes the demographic and background questions (Q2-Q13), Tables~\ref{questP2}~--~\ref{questP4} show the questions addressing readers (Q14-Q24) and all participants (Q25-Q30, Q44-Q51), and Tables~\ref{questP5},~\ref{questP6} include questions addressing only writers (Q31-Q39, Q40-Q43).

The survey design drew on previous studies \cite{AO3DemographicsSurvey2024, Rouse2023Overflow}, and was an iterative process. We conducted a small-scale pilot test with members of the fanfiction community, gathering feedback on clarity, comprehensiveness, and alignment with community norms. This process allowed us to refine the language and flow of the questionnaire. Feedback from the pilot also helped us identify gaps, such as underrepresented tools or platforms, which were subsequently addressed.
 


 We distributed the questionnaire to various fanfiction communities via email lists, Dreamwidth communities, private Discord servers dedicated to various fanfiction conventions, and social media posts. We also asked community members to distribute the questionnaire further. The study was conducted 
during May-July 2024.

\subsubsection{Data cleaning and consistency}

The original survey data included 489 respondents who identified as either fanfiction readers or writers, were adults, and signed the consent form. We then cleaned the data as follows. 
Using a geo-location service, we identified a consecutive bot attack originating from a single location (longitude, latitude), corresponding to 221 respondents who were removed from the dataset. We characterized the bot's behavior as answering only the minimal 44 required questions, each within 5-6 seconds. We then searched for additional respondents matching these characteristics and identified 111 more suspected bots. Further validation showed that these respondents answered no open-ended questions, and their email addresses were from servers that do not require authentication. Therefore, these 111 participants were also excluded.

The final dataset contains 157 respondents. Cronbach's Alpha was calculated for Q27 (1-8) (see Figure~\ref{fig:q27}) to assess the internal consistency of the multi-statement questions, yielding an average value of 0.85 with a low standard deviation. This high reliability indicates that participants responded consistently across the constructs, reflecting not only the robustness of the questionnaire design but also the attentiveness of respondents. The low variability in consistency suggested no clear evidence of inattentive behavior. These results confirm the validity of the data and its suitability for further study.  

Thus, our final dataset has 157 respondents, of which 90 are fanfiction writers and 67 exclusive readers. One writer identified as a non-reader. 

\subsection{Data analysis}
We conducted quantitative data analysis in Python (version 3.12.4), SPSS (version 29.0.2.0 (20)), and R (version 4.4.1), using the Python libraries numpy, re, itertools, pandas,plotly, matplotlib, and scipy, and the R library dplyr.  
 All multiple-choice responses were coded numerically.
Group comparisons were conducted using the Mann-Whitney U Test and the Chi-square test.  Responses to open questions averaged 23.3 words per answer per participant.

We conducted a content analysis of the qualitative data (Q29, Q36, Q46). Initial codes were generated from a preliminary review of the data by two independent coders, which were then consolidated into broader categories based on their frequency of occurrence. These categories were further analyzed to identify overarching themes. Inter-coder reliability was strong, with agreement rates of 86.6\%, 99.3\%, and 86.3\% for questions Q29, Q36, and Q46, respectively, across 100\% of the data. Appendix A includes our code books.





\section{Participants' demographics and community engagement} 
Here, we describe the participants' demographic information and their community engagement, as reported, including the platforms they use, the fanfics they engage with (either by reading or writing), and the AI tools that the writers among them use. Corresponding question numbers, as can be found in Tables~\ref{questP1}~--~\ref{questP6}, are in parentheses.   
\subsection{Participants demographics}
\begin{description}
\item [Gender] (Q3) Out of the 157 participants, 65\% (102/157) identified as women, 16.67\% (26/157) identified as non-binary/third gender, 12.7\% (20/157) as men, 2.5\% preferred not to say (4/157), and five self-described as follows: 
Both female and genderqueer, none, gender non-conforming, Agender, Nonbinary man, Agender femme.
\item [Continent] Asia 22.93\%, Europe 6.37\%, North America 63.05\%, South America 1.28\%, Prefer Not to Say 6.37\% 
\item[Primary language](Q6) 73.89\% (116/157) indicated English as their primary language, 24\% (38/157) primarily speak eight other languages, including Chinese, French, German, Hebrew, Hindi, Japanese, Spanish, and Russian. 
\item [Education] (Q4) 55.4\% (87/157) of the participants have an undergraduate degree, 25.48\% (40/157) have a post-graduate degree, 14\% have secondary education (24/157), 6 have vocational or technical training (e.g., apprenticeships, trade schools), 2 are undergrad students, and one preferred not to say. 
\item [Age] (Q2) 42\% of the participants (66/157) were 18-24 years old, 39.5\% (62/157) were 25-34 years old, 15.3\% (24/157) were 35-44 years old, 3.1\% (5/157) were 45-54 years old.
\end{description}

\paragraph{Comparison to known attributes of the fanfiction community} While data on fanfiction writer and reader demographics is limited, we compared our results to two significant surveys: the 2013 AO3 Census \cite{AO3Census2013} and the newly released 2024 AO3 Census \cite{AO3DemographicsSurvey2024}. The 2013 Census has served as a foundational reference for researchers. The 2024 Census offers an updated snapshot of the AO3 user base. Archive of Our Own (AO3) is one of the largest and most influential fanfiction platforms, with over 68,500 fandoms, 7.5 million users, and 14 million works \cite{ao3}. 

 The comparison suggests that AO3 demographics and our study sample share broadly similar age distributions, with a majority of participants in the 18–34 age range. Women are the majority in both datasets, though the proportion in our study is slightly higher. Significant representation of nonbinary and diverse gender identities is reflected in both datasets, but the 2024 AO3 Census captures a broader spectrum due to extensive categorization, while our study follows the CHI community recommendations for asking about gender \cite{genderHCI}. The geographic distribution shows regional variances between the 2024 AO3 Census and our study. Both highlight North American dominance, but our study shows higher representation from Asia and lower from Europe. Finally, both studies show strong representation of English-speaking participants. Our study’s multilingual representation aligns with the 2024 AO3 Census. It is important to note that both studies were administered in English.

\subsection{Engagement with Fanfiction}

The participants were, in general, fanfiction veterans (Q7). 46.5\% (73/157) were engaged for more than 10 years in the fanfiction community,  31.21\% (49/157) were engaged between 6 and 10 years, and 21.6\% (34/157) between 1 and 5 years. Most participants (135/157, 86\%) describe themselves as either highly or moderately engaged in the community (Q9). Participants who were writers (Q31) were more likely to have a higher degree and be engaged longer with the community.

\subsubsection{Fandoms and platforms}
\begin{figure*}[htbp]
    \centering
    \subfloat[Bar chart showing the top fandoms participants reported engaging with, along with the percentage of popularity among our participants (Q12)]{%
        \includegraphics[width=0.48\textwidth]{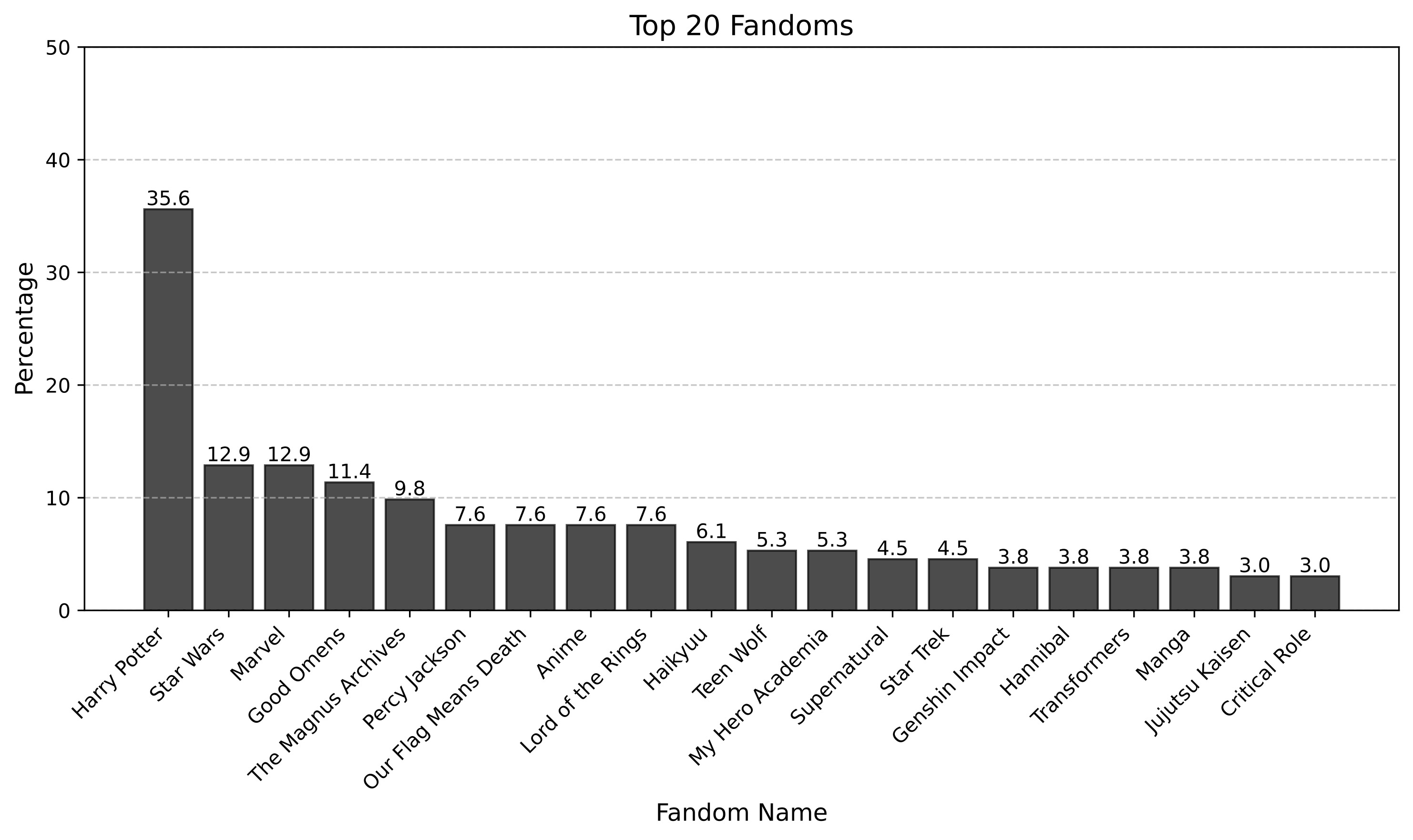}%
        \label{fig:fandoms}%
    }\hfill
    \subfloat[Bar chart showing the fanfiction platforms used by our participants (Q8)]{%
        \includegraphics[width=0.48\textwidth]{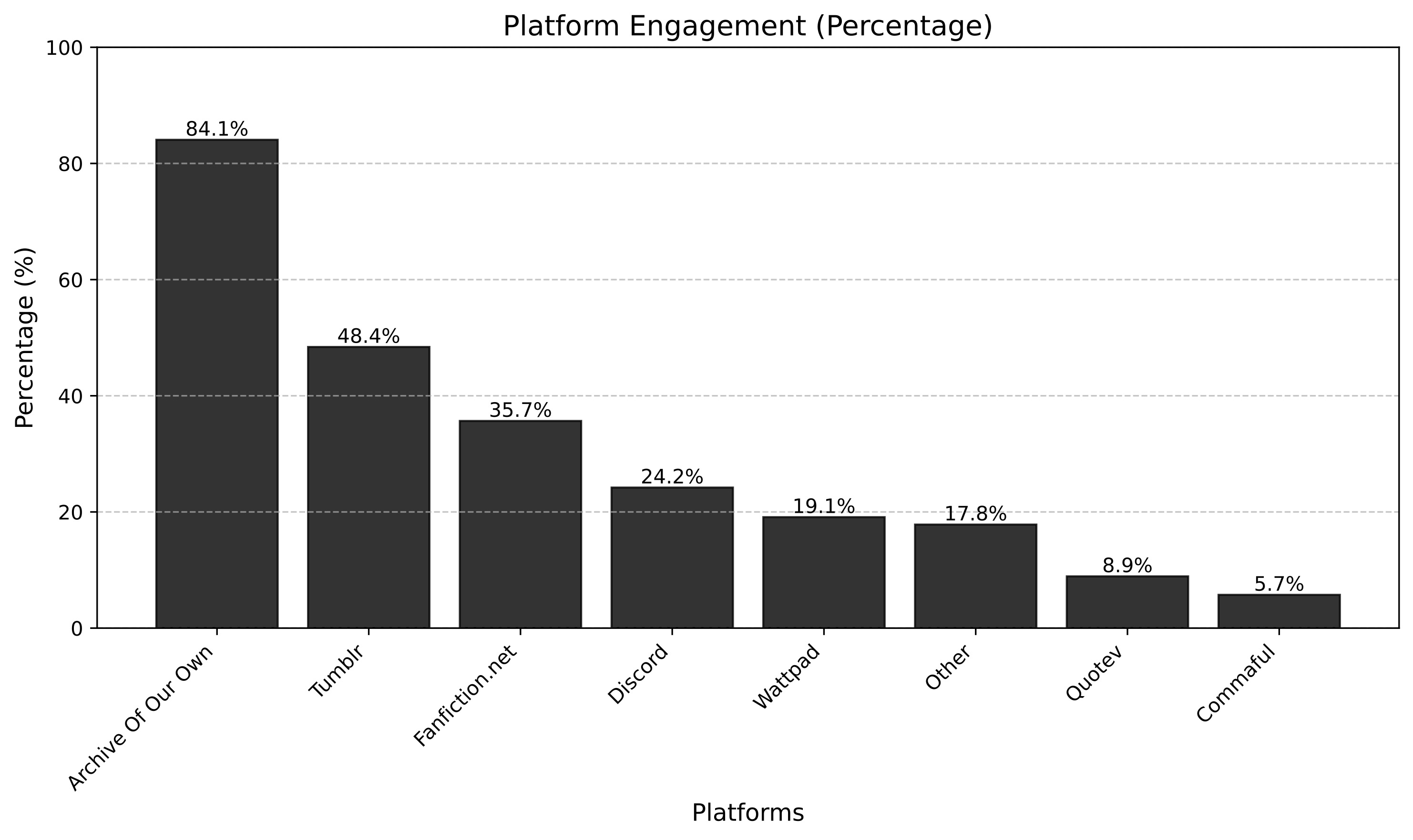}%
        \label{fig:q8}%
    }
    \caption{Prominent fandoms (a) and platforms (b) used by our participants.}
    \label{fig:side_by_side_figures}
    \Description{The figure consists of two images. On the left, labeled (a), is a Bar chart showing the top fandoms participants reported engaging with, along with the percentage of popularity among our participants. ``Harry Potter'' is the largest with 35.6\%, followed by ``Star Wars'' with 12.9\%. Other fandoms appear in decreasing order, representing varying popularity levels. On the right, labeled (b), is a bar chart showing the fanfiction platforms used by our participants. Archive of our own is the most popular, with 84.1\% of participants reporting using it.}
\end{figure*}

Our participants are engaged with 292 different fandoms, with Harry Potter being the most popular, followed by Star Wars in a distant second place. A bar chart showing the top fandoms participants reported engaging with, along with the percentage of popularity among our participants, is presented in Figure~\ref{fig:fandoms}. 

Overall, participants reported using a total of nine different platforms, with most of the participants engaging with the fandom using multiple platforms. AO3 is the most dominant platform. Figure~\ref{fig:q8} depicts a bar chart of the fanfiction platforms used by our participants. Archive of our own is the most popular, with 84.1\% of participants reporting using it.. 
  
\subsubsection{Digital tools in the writing process} 
  Writers were proficient in using prevalent online tools such as editors and search engines (Q33). 10\% (9/90) of the participants reported they primarily stick to pen and paper.

  We asked the writers which digital tools they use as part of their creative process (Q33). Most writers (84.4\% 76/90) reported that they perform online searches to assist with story development.  41.1\% (37/90) use digital note-taking and brainstorming apps. 11.1\% (10/90) reported the use of AI-powered tools such as Grammarly and ChatGPT.

\begin{table*}[h!]
\centering
\caption{Familiarity with AI Writing Tools (Q40\_1 -- Q40\_10). Use cases are typical uses as indicated by the tools' creators.}
\begin{tabular}{|>{\raggedright\arraybackslash}m{2cm}|>{\centering\arraybackslash}m{2cm}|>{\centering\arraybackslash}m{2cm}|>{\raggedright\arraybackslash}m{4cm}|}
\hline
\textbf{Tool} & \textbf{Familiar (\%)} & \textbf{Used (\%)} & \textbf{Use Cases} \\ 
\hline
ChatGPT & 89 (98.9\%) & 35 (38.88\%) & AI Assistant \\ 
Grammarly & 87 (96.66\%) & 42 (46.67\%) & Grammar, Proofreading \\ 
Jasper & 16 (17.97\%) & 7 (7.86\%) & Marketing, Copywriting \\ 
QuillBot & 25 (28\%) & 11 (12.3\%) & Paraphrasing, Rewriting \\ 
Wordtune & 21 (23.6\%) & 8 (9\%) &  Paraphrasing, Rewriting \\ 
Claude & 25 (28\%) & 11 (12.3\%) & AI Assistant \\ 
NovelCrafter & 21 (23\%) & 11 (12\%) & Novel Writing Assistance \\ 
AutoCrit & 17 (19.3\%) & 9 (10.2\%) & Editing, Style Assistance \\ 
SudoWrite & 34 (38.2\%) & 10 (11.2\%) & Creative Writing Assistance \\ 
Gemini & 48 (53.93\%) & 8 (9\%) & AI Assistant\\ 
\hline
\end{tabular}
\label{aitools}
\end{table*}

The vast majority of the writers are familiar with AI writing tools (Q40\_1 -- Q40\_10), and many are using them, as is detailed in Table~\ref{aitools}. Interestingly, 14.4\% (13/90) of the writers were familiar with all ten AI-based writing tools indicated in the survey. When asked about use rather than familiarity, 74.4\% (67/90) reported using at least one of the AI tools, and 33\% (30/90) reported never using any AI tools. A few indicated using most of the indicated ten AI tools.

\omitit{
}
\subsubsection{Engagement motivation}
Our participants joined the fanfic community (Q15) mainly to explore non-canonical relationships, which were not fully developed in the original story (68\%, 106/156) and for the opportunity to spend more time with their favorite characters (64.74\%, 101/159). 
Their reading engagement (Q16) is driven by prioritizing stories that allow them to further explore specific relationships and pairings (75.6\%, 118/156), as well as recommendations from trusted community guidance (64.1\%, 100/156). 
 
 The writers among our participants (90/156) explained their writing motivation (Q32) as driven mainly by engaging with a fandom: they were interested in character exploration 86.67\% (78/90), exploration of \textit{non-canonical} relationships (76.66\%, 69/90), and expanding the world of a specific fiction (61\%, 55/90). Engaging with other fans (56.6\%, 51/90) and easier publishing than in the industry (43\%, 39/90) were also highlighted. 
 Interestingly, the majority indicated that one of their reasons to write is to develop their writing skills (61\%, 55/90).

\section{Results: perceptions of fanfiction in the age of AI}

The survey's participants described fanfiction as a social community, with most participants viewing both the reading (74.5\%, 117/157) and writing (73\%, 115/157) of fanfiction stories as a social, community-centric activity (Q48\_3 , Q48\_4). The vast majority of participants (92\%, 145/157, Q48\_7) agreed or strongly agreed with the assertion  \textit{``Fanfiction is a space for human creativity''.}

\subsection{Perceptions of fanfiction authored with AI}
When asked about how accurately they could identify a story written with GenAI, the majority of participants (66.2\%, 104/157, Q49) were not sure. Yet, when asked whether they had previously read a story generated by GenAI, 57.69\% (90/156, Q19) of participants answered categorically that they hadn't, and only 16\% (25/156, Q19) were unsure.

Our participants did not believe that AI-generated content could replicate the emotional nuances and depth that are in human-authored stories (84.7\%, 133/157, Q48\_1). Neither did they think that AI would be able to maintain the authenticity of fanfiction narratives while offering innovative storytelling (77.5\%, 121/156 Q45).

We further asked which aspects they consider important when assessing a story written by AI (Q25), including in our analysis 63 participants who indicated they read or might have read a story generated by AI (Q19). Creativity and originality (54\%, 34/63), plot coherence (52.4\%, 33/63) and emotional depth (52.4\%, 33/63) were followed by writing style and fluency (36.5\%, 23/63), authenticity (33.3\%, 21/63) and character development (25.4\%, 16/63). 

\subsubsection{Readers' view on AI usage in the writing process} While the vast majority indicated fanfiction was a space for human creativity, when asked specifically about various uses of AI during the creative process, opinions varied. 

\begin{figure*}[!h] 
    \centering
    \includegraphics[width=\columnwidth]{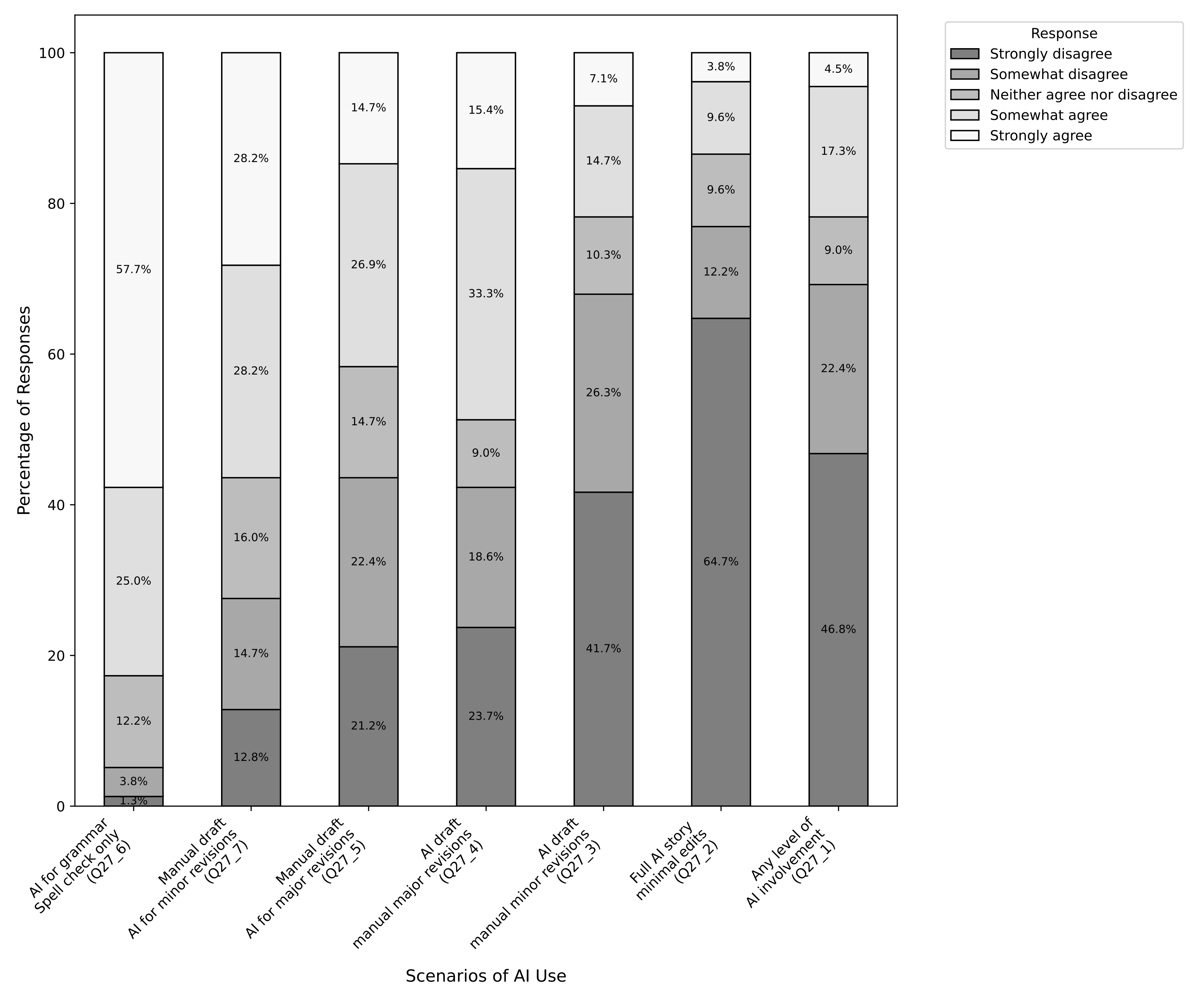} 
    \caption{Participants' willingness to read or engage with stories varies based on the level of AI involvement. A dark shade indicates participants who disagree to read stories with the specified level of involvement, a light shade indicates those who agree, and a medium shade represents those who are neutral.} 
    \label{fig:q27} 
\end{figure*}

We asked participants about their attitudes toward reading or engaging with stories that were created with various levels of AI involvement (Q27\_1 -- Q27\_8).  
The results, shown in Figure~\ref{fig:q27}, reveal conflicting views. While there is a general tendency to avoid stories with any AI involvement, participants express mixed opinions on various levels of AI use in drafting. Still, a vast majority (82.7\%) would read stories where AI was used solely for spell checking and grammar correction. 

When asked about their feelings regarding the increased integration of AI into fanfiction writing (Q44), only two participants expressed excitement. 26\% (41/156) of participants were cautious but open-minded, 24.36\% (38/156) indicated that their reaction would depend on how AI is implemented, and 3.85\% (6/156) were unsure. A significant proportion, 43\% (67/156), opposed the increased integration of AI. Interestingly, 83.58\% (56/67) of those opposing were also writers. Based on these findings, we further evaluate writers' specific views.

\subsubsection{Writers' views on AI use in their creative process}
\begin{figure*}[!h] 
    \centering
    \includegraphics[width=\columnwidth]{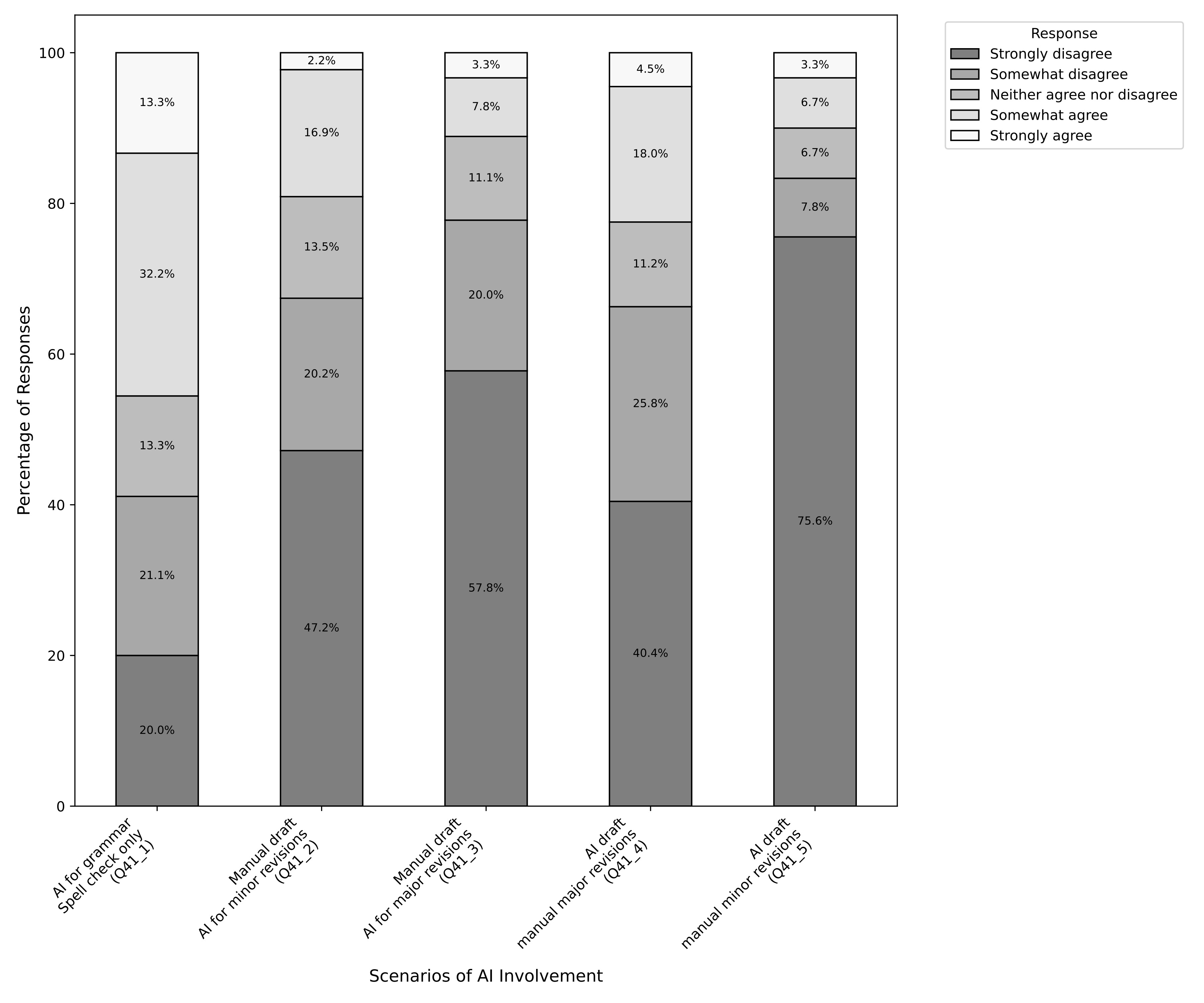} 
    \caption{Writers' views on incorporating AI into their creative process (Q41). The dark shade represents participants who disagree to read stories with the indicated level of AI involvement, the light shade represents those who agree, and the medium shade represents those who were neutral.} 
    \label{fig:q41} 
\end{figure*}
 
We asked writers (Q41) about their views on AI involvement in writing, phrased as: ``To what extent do you agree with the following statement: It is OK to...''. Figure~\ref{fig:q41} shows the results. Writers largely opposed AI creating the first draft or making major revisions to human-written drafts. 
Interestingly, 10\% (9/90) of the writers supported the idea of AI fully generating fanfiction stories. We further found that these supporters were all highly familiar with various AI tools and reported they used some of the tools ``a lot'' (Q40).

When asked about how they view the role of AI in their own writing process (Q39), 40.9\% (36/90) of the writers did not find it played any role, and 27.7\% (25/90) viewed it as an assistant or aid for specific aspects of the writing. 
Only 5.5\% (5/90) viewed AI tools as either co-creators or creative helpers. Interestingly, more than 17\%  (15/90) viewed AI tools as a competition (``A tool that other people use that I have to compete with'').

\begin{figure*}[htbp]
    \centering
    \subfloat[Writers' perspectives on acceptable scenarios for collaborating with AI (Q42).]{%
        \includegraphics[width=0.7\textwidth]{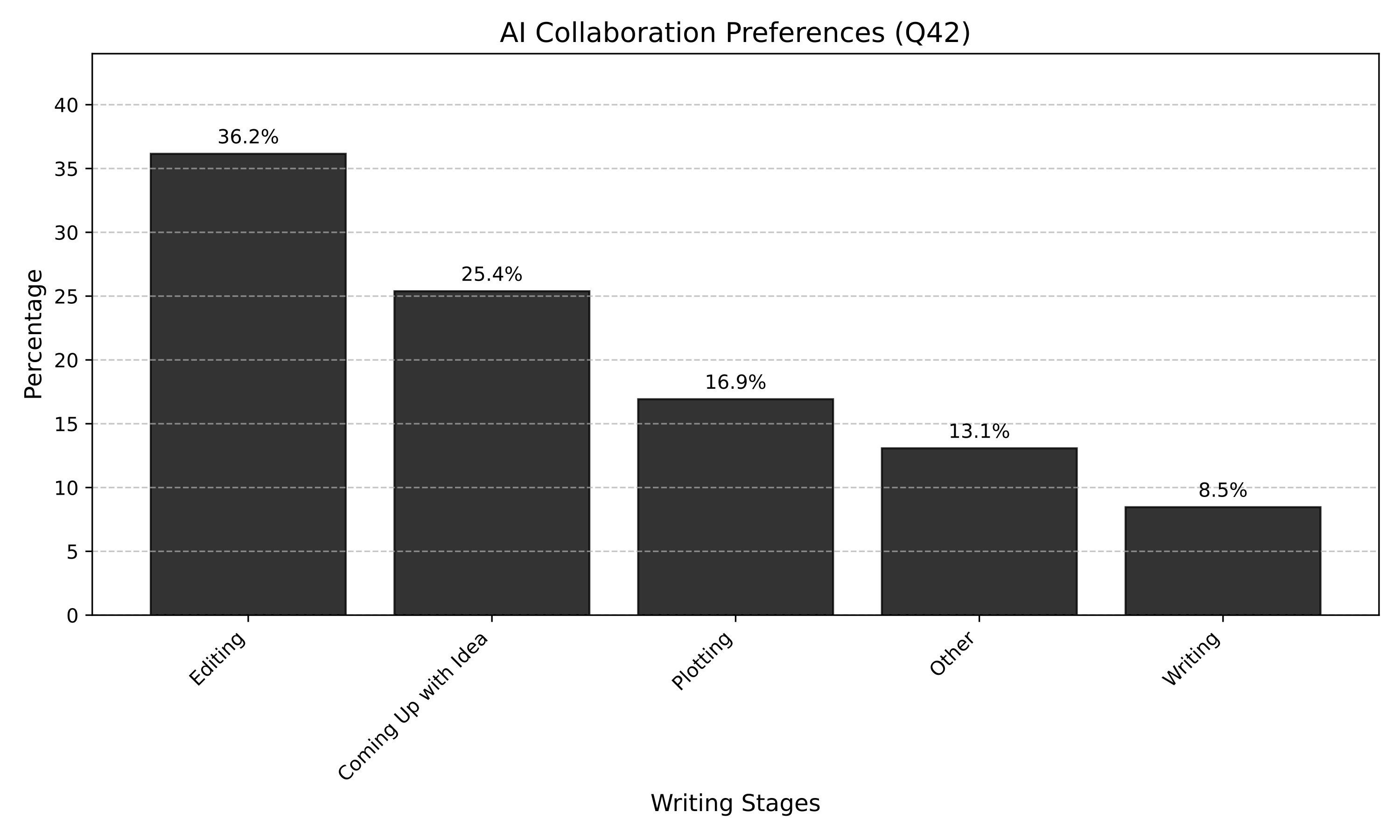}%
        \label{fig:q42}%
    }
    \vspace{0.5cm}
    \subfloat[Writers' perspectives on the aspects of writing where AI would be most beneficial (Q43).]{%
        \includegraphics[width=0.7\textwidth]{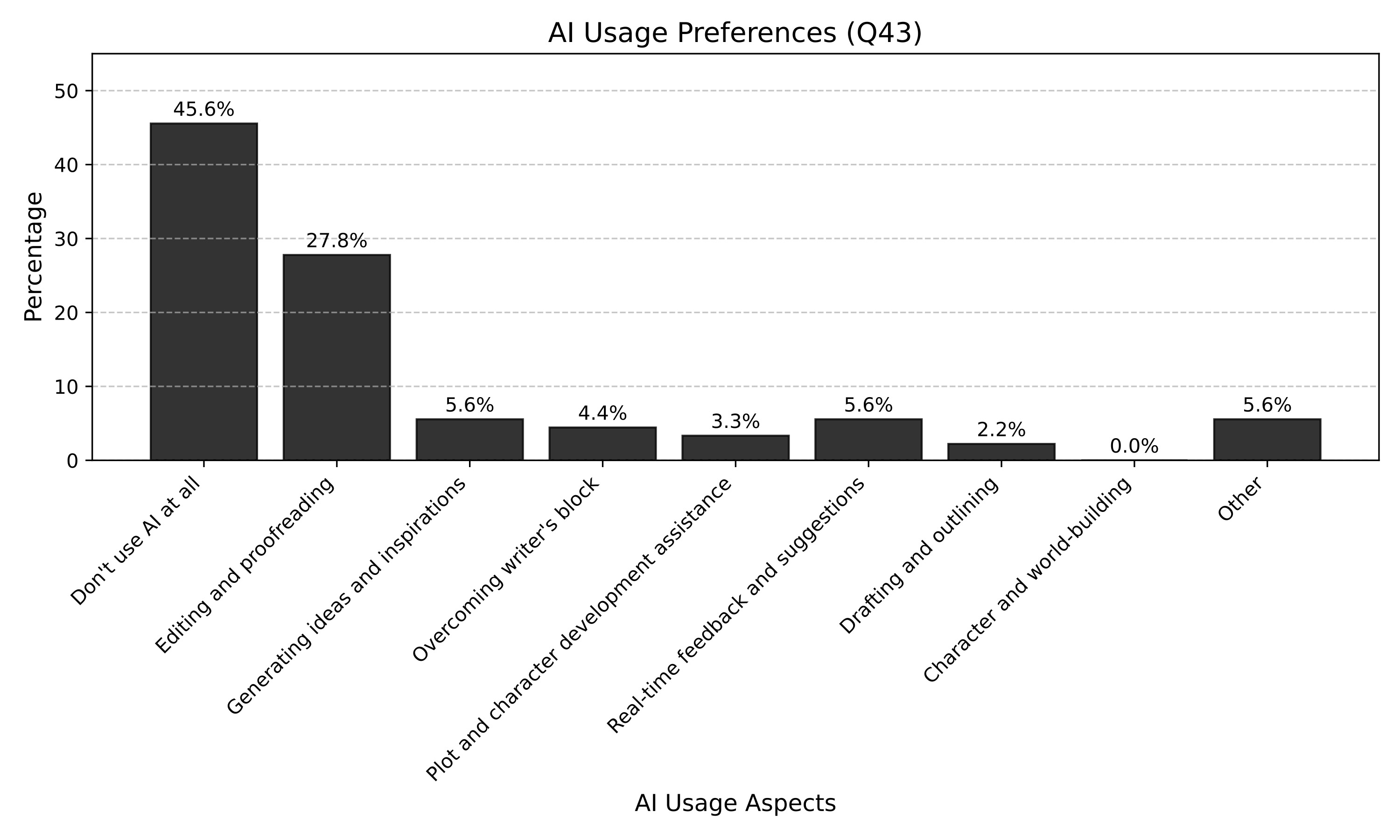}%
        \label{fig:q43}%
    }
    \caption{Writers' perceptions regarding (a) acceptable scenarios for collaborating with AI and (b) aspects of writing where AI would be most beneficial.}
    \Description{The figure consists of two images. The top image, labeled (a), is a bar chart showing Writers' perspectives on acceptable scenarios for collaborating with AI (Q42). The bottom image, labeled (b), shows Writers' perspectives on the aspects of writing where AI would be most beneficial (Q43).}
     \label{fig:q4243}
\end{figure*}

Figure~\ref{fig:q4243} depicts the writers' answers regarding their perceptions on acceptable scenarios for collaborating with AI (Q42) and aspects of writing where AI would be most beneficial (Q43). 
Most writers the use of AI for editing acceptable (36\%, /72) and 65\%  (47/72) found it acceptable for the AI to generate story ideas.  30.56\% (22/72) were comfortable with consulting the AI during the plotting phase. 

Yet, when asked about their writing process (Q43), many writers (45.5\% 41/90) indicated they do not use AI at all. Of the remaining participants, 27.7\% (25/90) found AI beneficial for editing and proofreading, with an interesting finding that 65\% (16/25) of them indicated \textit{English} as their primary language.  

\paragraph{Writers' predicted impact of AI. }
In response to an open question about hopes and concerns regarding AI and writing (Q46), several writers highlighted benefits of AI writing tools, such as increased engagement ``If used properly, it can help more fanfic writers publish quicker and actually finish stories'' (P36, more a reader than a writer). Others expressed concerns 
that writers would be replaced or pushed out: ``Biggest concern that young girls will stop writing because they can't compete and the ai is biased aGenAInst them'' (P25, reads more than writes). 


\subsubsection{
Differences in writers' views of AI based on experience levels}
The high variance in writers' responses motivated us to analyze their answers based on their experience within the community, as indicated by their responses to Q7 (duration of involvement in fanfiction).
Our analysis revealed significant differences in perceptions of AI involvement in the writing process among writers with varying levels of experience.

We analyzed Q41, which included five statements about AI involvement in the writing process, where participants indicated their level of agreement, and Q43, where they identified the writing aspects they believed AI would be most beneficial for.

Analyzing Q41 across experience levels, we find that most participants, across experience levels,  generally tolerated limited AI  assistance. 
As more AI involvement was suggested, answers varied across the groups more.  
  Initially, when AI played only a minor role (Q41\_1), writers generally agreed on its acceptability, with no significant differences among experience groups (e.g., 1–5 years vs. 6–10 years: U=197.0, p=0.083; 6–10 years vs. >10 years: U=559.0, p=0.979). However, as AI involvement increased, a clear pattern emerged. Even at moderate AI usage (Q41\_2), the most experienced writers (>10 years) were significantly more resistant than the newest writers (1–5 years: U=492.5, p=0.0448). This gap increased at higher levels of AI involvement: for example, at Q41\_3 (U=515.0, p=0.013), Q41\_4 (U=502.0, p=0.0355), and Q41\_5 (U=566.0, p<0.001), early-career writers (1–5 years) consistently showed greater acceptance than their veteran counterparts. Writers with 6–10 years of experience tended to occupy an intermediate position, sometimes differing from 1–5 year writers (e.g., Q41\_5: U=211.5, p=0.038) but not significantly diverging from those with more than 10 years. 
Thus, early-career writers generally embraced AI assistance more readily than the others.

A chi-square test of independence was conducted to explore the relationship between writers' duration of involvement in the fanfiction community (Q7) and their perceptions of AI's role in writing (Q43). Significant differences were observed between groups. Writers with 1--5 years of experience were more likely to embrace creative AI applications, such as generating ideas or assisting with plot and character development (\(\chi^2(7) = 17.581, p = 0.014\)). In contrast, writers with over 10 years of experience overwhelmingly favored avoiding AI altogether, with "Don't use AI at all" selected far more often than expected. Similarly, significant differences were found between writers with 6--10 years and those with more than 10 years of experience (\(\chi^2(7) = 15.441, p = 0.031\)). Writers in the 6--10 year group leaned towards technical applications, particularly editing and proofreading, while those with over 10 years of experience aGenAIn demonstrated a strong preference aGenAInst AI use. 

Overall, we see a trend of resistance to AI among more experienced writers, while newer and moderately experienced writers are more open to integrating AI for both creative and technical tasks. This suggests a potential cultural shift in attitudes toward AI within the fanfiction community.

\subsection{Concerns and divides: perspectives on AI's impact on creativity, quality, and ethics in fanfiction}
When asked about the emergence of AI-authored fanfiction (48\_5, 48\_6, 48\_8, 48\_9), 76.4\% (120/156) thought that it poses a danger to the social aspects of the community (Q48\_5), and 83.4\% (131/157) were concerned that an influx of AI-created stories might inundate fanfiction platforms, potentially overshadowing human-authored stories (Q48\_6). There was a strong concern for human creativity, as 79.6\%  (125/157) asserted that relying on AI for content creation could stifle human creativity and creative freedom (Q48\_8). 
Conversely, 31\% (49/157) of participants thought that worries surrounding AI's role in fanfiction might be exaggerated (Q48\_9). 

\subsubsection{A space for human creativity}

We asked participants what concerns, if any, they have about using AI in fanfiction. First, participants selected from a list (Q51) (Loss of personal touch; Over-reliance on technology; Copyright infringement; Inaccurate character portrayal; Ethical implications; Low-quality content; Impact on emerging writers; Loss of personal journey; Other), and then elaborated in an open question (Q36). 

The main emerging concern was that the community would lose its special place as a space for human creativity. 
Participants worried that the use of AI would result in a loss of personal touch or authenticity (78.2\%, 122/156, Q51). For example, in the words of one participant (P95, both reader and writer, Q36): ``AI is the death of human creativity. Instead of exploring their own imagination, AI lets you just 'create' mindlessly.'' 

A majority of participants were concerned about overcrowding spaces with low-quality AI-generated content (73.7\%, 115/156, Q51). Quoting some participants: ``They really feel low quality mostly and I wouldn't want to be scrolling for hours without finding a fic that I can actually read without it feeling so off''  (P15, reader more than a writer, Q36).

Many (67.95\%, 106/156) also thought that GenAI would fail to capture the essence of the fandom community and its characteristics correctly. For example, P4 (both reader and writer, Q36) shares, ``It seems cold, detached, and cynical to produce millions of stories by AI. We are in the fandoms because of the warmth, I don't need to feel like a machine is cuddling me. It'll probably starve my soul without me even realizing it''. Other participants mentioned appreciation of fans' labor and time invested in creating a story, as well as human interest and passion for a fandom.

Alternatively, 51.9\% (81/156) were concerned that the use of GenAI tools would lead to a loss of the writing experience as a personal journey. In the words of P50 (who reads more than writes, Q36)

``...I feel strongly that the point of something like fanfiction, especially, is not just to generate a text, but to learn something in the process of doing so, whether that's a greater understanding of the source material or simply how to write better; there's also the engaging-with-the-fan-community factor'.'

Low quality was also a big concern. When asked about their biggest hope or concern regarding the intersection of AI and fanfiction (Q46), 
15.5\% (14/90) responded that their concern is that AI makes low-quality stories - ``Stories will become worse and less inventive'' (P116, writer).
There were concerns about the future of the space as a result.  
``a flood of poor-quality, repetitive ffns and a loss of motivation from the human writers'' (P158, read more than writes).

\subsubsection{Ethics. } Ethics emerged as an additional significant concern, as indicated by 68.6\% (107/156) of participants (Q51). Half of the participants expressed concerns about potential infringement and originality. For example, one participant wrote ``I don't need auto-plagiarism involved in my work, or to have the most statistically obvious wording.'' (P109, both reader and writer, Q36). Other participants expressed concerns regarding the legal aspects of copyright law and

Additional participants voiced concerns about using content created by community members for training AI without consent from authors: ``It's theft from actual writers'' (P143, reader, Q36). 
A few participants, 6.4\% (10/156), also indicated concerns about data privacy and AI environmental impact, with 7.6\% (7/91) of the responses expressing such concern.

\subsection{The perceived effect of overt and covert AI agents on the community} 

When asked about whether knowing a fanfiction story was generated by AI would decrease their interest in reading it (Q22), most respondents (66\%, 103/156) indicated that it would, while a smaller group (16\%, 25/156) stated that their interest would depend on the context. Only a minority expressed increased interest (5.77\%, 9/156) or no influence (3.85\%, 6/156).

Participants considered transparency as a crucial aspect of using AI in creative writing, with 86\% (135/157, Q48\_2) asserting that authors should be transparent about the involvement of AI in their creative writing process. When asked how they would feel if they found in retrospect that a fanfiction they read had been written by AI (Q21), 72.2\% (112/155) described negative feelings about it, with 58\% (90/155) feeling deceived. Yet, 9\% (14/155) said they would be impressed by such a revelation, describing a positive experience, and an additional 5\% (8/155) were agnostic. 

We then asked whether the notion that a fic might have been created using
GenAI influenced our participants' interactions within the fandom community, and if so, how (Q29). More than half of the participants responded with no influence (54.3\%, 56/103). In the words of one participant ``Artificial intelligence or not, as long as the content is good'' (P75, reader, Q29). One participant mentioned positive influence: ``It depends, I've been apart of 'dead' fandoms where there has not been any recent fanfiction updates in months/years and would like to have some sort of new content to read'' (P36, more reader than a writer, Q29).
38.8\% (40/103) of the participants mentioned moderate or strong negative influence. Of these, 22.5\% (9/40) participants mentioned actively avoiding AI-generated fiction, 27.5\% (11/40) general disdain of AI, 17.5\% (7/40) pointed to its limited creativity, and 20\% (8/40) view AI creations as plagiarized or inauthentic works. For example, one participant wrote ``I do not want to read AI-generated fic or engage with people who post AI-generated fic, especially on platforms focused on fic, such as AO3. If I encounter someone on AO3 who admits to generating their fic with AI, I block them and hide all works by them...'' Some other participants also mentioned blocking fans who use AI for generating fics.  

When asked about their biggest concern or hope regarding the intersection of AI and fanfiction, participants often referenced the future of the community and the connections formed within it (Q46). A few expressed optimism and hopes for the community. For example, P86, reader: ``Hope: AI will foster a culture of collaboration and mentorship within fanfiction communities, encouraging knowledge sharing and skill development''.  Many others expressed concerns, including potential devaluing of fans' labor, and interest in specific fandoms rising and falling rapidly based on AI-driven activity, potentially undermining the depth and sustainability of fan engagement. A major concern is discouraging interactions between fans. In the words of one participant: 
 
``One of the things I love most about fanfic is the community, to know that someone feels the same about a character that I love or hate or have a different opinion for an interesting reason. Ai can't feel, can't have informed opinions'' (P17, a reader).

\omitit{
\begin{figure}[!h] 
    \centering
    \includegraphics[width=.7\columnwidth]{quality_assessment_radar_with_legend.pdf} 
    \caption{Prominent aspects when evaluating the quality of a story, whether human-created (Q13) or AI-generated (Q25), include emotional depth, plot coherence, originality, and writing style. Expectations for story quality remain consistent regardless of the creation method. Emotional depth and plot coherence are most valued (79.6\%, 125/157, and 50.3\%, 70/139), followed closely by originality and writing style (63\%, 99/157, and 43\%, 60/139, respectively, Q25).} 
    \label{fig:quality} 
  
\end{figure}
}

\section{Discussion}
Creative communities are undergoing significant disruption, with the long-term outcomes still uncertain~\cite{hbr2023bgenerative}. 
To understand this transition phase, its dynamics, and how it is perceived within these communities, we turned to fanfiction communities.  
We present findings from a study involving 157 members of fanfiction communities, who are fanfic writers and/or readers. We explored participants’ views on the integration of GenAI in their space, their
knowledge of and engagement with GenAI tools and GenAI-generated content, their motivations for incorporating (or rejecting) AI in their creative processes, and their perceptions of AI’s role in shaping the future of storytelling within fanfiction communities. Our study contributes to the growing body of research that examines the impact of GenAI on creative practices (e.g., \cite{hbr2023generative, shumailov2024ai}), and extends it by investigating the unique dynamics of \textit{participatory} creative practices within fanfiction communities. Fanfiction communities are defined by their participatory, collaborative, and iterative practices. Following, we discuss our key findings and their implication for online fanfiction spaces. 
\subsection{RQ1: Understanding the Characteristics of the Fanfiction Members Study’s Participants}
The participants in the study were diverse, both in demographic backgrounds and fanfiction experiences. A majority of the participants were women, with a notable presence of non-binary individuals, reflecting the inclusive nature of fanfiction communities. This gender distribution is consistent with both recent and historical demographic surveys of fanfiction communities \cite {AO3DemographicsSurvey2024, Rouse2023Overflow, AO3Census2013, BaconSmith1992EnterprisingWomen}. 
Our study participants are characterized by different levels of engagement with a wide variety of fandoms. Their engagement varies from casual readers to dedicated readers and writers. Many of the participants have been active in fanfiction for over a decade. Most of the participants view fanfiction as a creative and social space where collaboration and feedback from other writers and readers are highly valued. This view resonates with Jenkins' description of ~\etal~\cite{jenkins2016participatory} participatory culture.  The goal of RQ1 - Understanding participants' characteristics as members of fanfiction communities is critical for contextualizing our findings about the changing dynamics and forces within the community, and their perception and adoption of GenAI tools and generated content.

\subsection{RQ2: How Do Fanfiction Community Members Perceive Human Versus AI-Generated or Collaborative Content, and How Do These Perceptions Influence Community Dynamics?}

\subsubsection{Quality of AI-generated content}
While the majority of participants were not sure they could accurately identify a story written with GenAI, our findings suggest that fanfiction readers and writers overwhelmingly favor human-created works, valuing the emotional authenticity, character consistency, and creativity that human authors bring to fanfiction. Participants perceive GenAI-generated fics as low quality in terms of their limited emotional depth, complexity, and authenticity, as a major factor in their rejection of AI-created stories. It is important to point out an inconsistency in the participant responses - the perception of the AI-generated content as low quality is prominent despite the fact that participants' responses indicate that they were not sure they could accurately identify a story written with GenAI. 

Rapid advancements in LLMs and GenAI-powered writing assistants are likely to influence the acceptance of GenAI-generated stories.  Ethan Mollick, in his book Co-Intelligence, encourages users to \textit{``assume this is the worst AI you will ever use''} \cite{mollick2023cointelligence}, suggesting that future iterations of GenAI tools will address many of their current limitations. As models evolve, personalized writing assistants with ``Write Like Me'' functionality - trained on a user’s own writing to mimic their tone, voice, and style — may produce content that feels more authentic and closer to human-generated stories. Such technological improvements could possibly lead to increased acceptance of GenAI-generated stories within fanfiction communities.

\subsubsection{GenAI transparency and its implications for writers and communities}

Participants expressed concerns about the ethical implications of using AI to create stories,  particularly regarding the intellectual property of human authors whose works are used, often without consent, to train AI models. This aligns with ethical debates raised by Sittenfeld \cite{sittenfeld2024experiment} and Frenkel \cite{hbr2023generative}, who caution that using copyrighted material without explicit consent poses significant challenges to the fair use of AI in creative fields. Similarly, recent research indicated that creative writers express concerns about the use of their work in training AI models, emphasizing the importance of maintaining control over their creations~\cite{gero2024creative}. 
This concern highlights a need for transparency regarding the data used for training LLMs. In particular, our participants, who are mostly non-expert users of GenAI tools, seek to understand what datasets were used to train the models, which were in turn used to create stories, and how these datasets were constructed and curated. 

Participants also advocated for greater transparency when AI is involved in content creation, suggesting that readers should be made aware of whether a work was AI-generated or when AI is involved in content creation. This call for transparency is not unique to fanfiction communities but echoes concerns in other areas, including education \cite{khan2024brave}, legal writing \cite{shope2023best}, scientific writing \cite{harker2023authorshipguidelines}, and political advertising \cite{carr2024ai_political_ads}.

A recent article by Liao and Vaughan \cite{Liao2024AI}, advocates for the development of novel human-centered strategies to enhance transparency in LLMs, aiming to address the needs of diverse stake holders. Our findings highlight the importance and urgency of applying such an approach.

Transparency has the potential to transform not only reader trust but also the mechanisms through which stories are recommended. For instance, could knowing whether a story was AI-assisted impact how content is recommended? Would transparent labeling discourage recommendations for AI-heavy works, thus preserving human-centered storytelling? We discuss this issue further in the section on Implications for Design.

\subsubsection{Preserving Community Values}

Community members expressed various opinions on the future of the community, revealing changing social dynamics. Some were intrigued by the possibilities of having AI voices in the community, and others suggested, similar to~\cite{hbr2023generative}, that the community would become flooded with low-quality content. 
Other participants expressed fundamental concerns over GenAI-generated fics undermining fanfiction's core values, including human creativity, originality, and community engagement. Similar concerns were raised by Li and Pang \cite{li2024fandom}, who discuss how AI-driven content threatens the participatory culture of fandoms by introducing GenAI-generated content into a creative space, which assigns high value to human labor and contribution. 

The fear of losing the human-centric nature of fanfiction communities contributes to participants' wariness of GenAI as a \textit{co-creator}, and explains its occasional view as \textit{competitor}. 

Our findings suggest that while these issues resonate with participants, their prevalence today remains nuanced. Many writers already use various technologies, such as Grammarly or brainstorming apps, to support their creative processes. Additionally, a subset of participants reported actively using AI tools like ChatGPT or Jasper. Does this widespread use of writing aids mean that the erosion of originality and autonomy has already begun?

Current literature provides mixed insights. On one hand, researchers highlight the homogenization risks posed by overreliance on AI \cite{Bender2021, brinkmann2023machine, doshi2024generative}, which can reinforce biases and reduce narrative diversity.  On the other hand, AI tools can serve as creative aids, offering suggestions that spark new ideas \cite{ShaerBrainwriting24, mollick2023cointelligence}. Future research should examine whether increased reliance on AI tools in writing correlates with declining originality in fanfiction specifically, and in storytelling more broadly, or whether AI simply augments existing creative practices. 

In addition to concerns about transparency in the generation of content, participants raised important ethical questions about the datasets used to train GenAI models. Specifically, we found apprehension that fan-created works—often produced within a gift economy and shared freely within communities—may be scraped and repurposed for training commercial AI systems without consent or attribution. This finding raises fundamental issues of ownership, intellectual labor, and community norms, particularly in a context where fanfiction is not only creative but deeply personal and communal. The unauthorized inclusion of such works in training corpora challenges the ethos of fan-driven spaces and may be perceived as a violation of both legal and moral boundaries. This finding echos broader debates in AI ethics around data provenance, informed consent, and fair compensation for creative labor. 

Recent investigations into creative communities have revealed widespread apprehension regarding the unauthorized use of artistic works in AI training~\cite{kyi2025governance,lovato2024foregrounding}. These studies highlight creators' frustration over the lack of acknowledgment, consent, and meaningful compensation when their content becomes part of massive training datasets. Efforts have been proposed to improve consent management and attribution transparency~\cite{balan2023decorait}, and legal actions in multiple jurisdictions are now seeking to establish clearer boundaries around the use of creative work in this context~\cite{apfrenchmeta2025}. 

A recent study of creative writers'
attitudes found that writers and creators express that the transformation of creative expression into ``data'' fundamentally alters the meaning and value of their work~\cite{gero2025creative}. The study showed that they are particularly concerned with the decontextualization of their voice, identity, and lived experience, especially when writing is deeply personal, autobiographical, or represents marginalized perspectives. The lack of agency in this process—combined with power asymmetries between creators and technology companies—further contributes to a sense of exploitation. Many creators emphasize that the intended audience of their work is human, not algorithmic, and they reject the notion that their labor can be treated as an undifferentiated input into computational systems~\cite{gero2025creative}. These perspectives frame the use of creative content in AI as not just a technical issue but, similar to our findings, an ethical one requiring deeper participatory deliberation and the development of systems that respect the humanity embedded in creative work.

Addressing these challenges will therefore require not only technical solutions (e.g., dataset transparency and opt-out mechanisms) but also community-informed governance approaches that respect the values of participatory creative cultures.

\subsubsection{Perceptions regarding Potential Loss of Expertise}
In that respect, some discussed the possible loss of expertise, suggesting that community members would miss out on the experience of developing their talent and craft with time and feedback.

These concerns are unlikely to be alleviated by increased adoption of GenAI tools in other writing communities, such as business~\cite{coman2024perceptions, hbr2023generative} and education~\cite{khan2024brave}, and have implications for the design of online fanfiction communities. 

Participants were united in their view of the community as a space for human creativity, and many highlighted the importance of the ``engaging-with-the-fan-community factor'' (P50), and as P4 shared:  
     ``We are in the fandoms because of the warmth, I don't need to feel like a machine is cuddling me.''

\subsection{RQ3: How are AI and Large Language Models perceived within the fanfiction community?}
Our findings reveal a nuanced view of AI in the fanfiction community. While some participants viewed AI-generated content as fundamentally incompatible with the values of fanfiction—particularly around authorship, community ethics, and emotional authenticity— others described it as a useful creative companion. These writers emphasized the pragmatic potential of AI to support idea generation, assist with world-building, or help navigate blocks in the writing process. For some, especially newer and self-publishing authors, GenAI was framed less as a threat to creativity and more as a tool to extend it. 
This view is aligned with previous studies that have shown AI and LLMs’ potential to assist in idea generation, narrative, and character development \cite{prewriting2024,ShaerBrainwriting24, girotra2023ideas, yang2022ai, redhead, plan-and-write, Zhao2024NarrativePlayAA}. However, many participants were skeptical about AI's ability to capture the emotional depth and creativity inherent to human-authored fanfiction. This skepticism is consistent with concerns raised in both scientific literature (e.g. \cite{doshi2024generative, Bender2021}) and popular media (e.g. \cite{sittenfeld2024experiment}) about the limitations of AI in writing nuanced, authentic, and emotionally deep stories.

These perspectives highlight that attitudes toward AI are not uniformly negative; rather, they reflect a spectrum shaped by experience, writing goals, and positionality within the community.
 
\subsubsection{Concerns of Experienced vs. New Writers}

Most participants in our study were engaged in fanfiction communities for more than six years. These participants expressed concerns about losing the personal journey of writing. However, our findings indicate that these concerns might not apply equally to newer or first-time writers.

Newer writers, especially those more familiar with digital tools, often express openness to AI as a creative aid. These participants are more likely to view AI as a source of inspiration, idea generation, or productivity boost, rather than a threat to authenticity. 

Research on digital creative learning environments suggests that newcomers often benefit from structured guidance and tools, which can include AI \cite{luckin2016intelligence, aragon2019writers, khan2024brave}. For instance, AI-powered systems could provide scaffolding for first-time writers, such as real-time feedback or grammar corrections. However, overreliance on these systems risks hindering skill development and critical engagement with the creative process. Further inquiry is needed to understand whether new writers experience the same anxieties about creative autonomy or whether they view AI as an enabling tool to enter the community.

This generational or experiential divide underscores the importance of designing AI tools that can accommodate multiple modes of creative engagement, supporting those who seek assistance while preserving space for interpersonal mentorship and community recognition.

\subsubsection{Adoption of GenAI Tools}
Our findings indicate that fanfiction writers with more experience using GenAI tools were more open to incorporating these technologies into their creative processes. For example, 10\% (9/90) of the writers, who also reported being highly familiar with various GenAI tools, supported the use of AI to fully generate fanfic stories (Q40). Results from a recent study on the ``perceptions of professionalism and authenticity in AI-assisted writing''~\cite{coman2024perceptions} similarly indicate that professionals who frequently use GenAI tools in their work are more likely to view AI-assisted writing as authentic, effective, and confidence-building.  These findings suggest that familiarity with GenAI, whether in professional or creative contexts, may lead to a greater acceptance of AI-generated content, as users become more familiar with its capabilities and limitations.

These small groups of early adopters are also aligned with Rogers' Diffusion of Innovation (DOI) theory \cite{rogers2003diffusion}, which claims that early adopters of new technologies tend to influence broader adoption within a community by showcasing its practical benefits. As early innovators in the fanfiction community become more proficient with GenAI tools, they may contribute to wider acceptance of these technologies. 

Interestingly, we find that novice writers were more open to incorporating AI at various levels of the writing process compared with more experienced writers. This can be seen through the lenses of forming an identity within the writers' domain, as writers with an established writing identity need to engage more in the sense-making of the change and its impact compared with novices~\cite{schmitt2024generative}.


\subsection{Theoretical Implications}

Our analysis is informed by the theoretical foundation outlined in Section~\ref{sec:theory}, drawing on participatory culture, creative labor, and AI ethics to frame our interpretation of participants' responses. These frameworks help illuminate how the adoption of generative AI is not experienced merely as a technical shift but as a disruption to deeply held community norms and creative practices. 

Our findings raise important theoretical questions about authorship, agency, and participation within digital creative environments shaped by generative AI. In communities like fanfiction, creative production is not simply an individual act but a socially embedded process involving attribution, collaboration, and shared meaning-making. These participatory norms stand in tension with the extractive logic of many AI systems, which are often trained on large-scale datasets assembled without the consent, credit, or awareness of the original content creators. This disjuncture calls for a closer theoretical interrogation of how AI technologies disrupt established understandings of authorship and the socio-technical contracts that underpin collective creative practices.

Drawing on Jenkins’s work on participatory culture~\cite{jenkins2006convergence,jenkins2008convergence}, we see fanfiction as a domain characterized by low barriers to entry, distributed authorship, and strong community norms of collaboration and feedback. However, when user-generated content is repurposed as training data for commercial AI models, these participatory dynamics are subverted. Rather than facilitating collaboration, generative systems may fragment and anonymize creative labor, introducing new risks of misattribution, cultural appropriation, and erasure~\cite{jenkins2008convergence,jenkins2016participatory}.

These concerns are not merely legal or technical. Within the fanfiction context, creative works are often imbued with emotional, cultural, and personal significance. They emerge from trust-based ecosystems of voluntary labor, where sharing is motivated by community recognition rather than commercial reward. The unconsented use of such work in AI training—especially in commercial applications—can lead to a sense of violation and loss of agency. Couldry and Mejias~\cite{couldry2019data} conceptualize this dynamic as data colonialism, wherein everyday cultural expression is commodified without reciprocity, further entrenching unequal structures of ownership and control.

Moreover, the assumption that data used in AI systems is neutral or context-free has been widely critiqued. Jo and Gebru~\cite{jo2020lessons} argue that data infrastructures embed social values, institutional biases, and historical inequalities. When creative outputs are removed from their cultural context and treated as mere linguistic inputs, the systems that reproduce them risk misrepresenting or flattening the intent of the original work. Gero et al.~\cite{gero2025creative} observe that fanfiction authors often experience this form of AI-driven repurposing as misrecognition or disempowerment, particularly when their stylistic voice is mimicked or reassembled without consent or attribution.

These ethical concerns are compounded by the asymmetrical power relationships that structure AI development and deployment. The boundaries between human and machine authorship have become increasingly blurred, and with that, the meaning of co-creation becomes more ambiguous. As Jo and Gebru~\cite{jo2020lessons} and Lovato et al.~\cite{lovato2024foregrounding} note, questions of agency, visibility, and ownership are not evenly distributed within these sociotechnical systems. A system that draws freely from community-generated content without involving those communities in design or governance reflects not co-creation, but a reconfiguration of power.

Rather than viewing generative AI as a neutral technological tool, it is more appropriate to conceptualize it as a cultural system embedded with human assumptions, institutional agendas, and structural asymmetries~\cite{seaver2017algorithms,brinkmann2023machine}. This framing allows us to interpret participant concerns not only in terms of functionality or misuse but in relation to the broader ethical and cultural logics inscribed in AI infrastructure.

Mentorship offers a concrete example of these shifts. In fanfiction communities, mentorship—through beta reading, peer feedback, and collaborative writing—has historically been central to learning and socialization~\cite{black2008adolescents,de2021rogue}. As generative tools become more common, they may replace or reconfigure these roles. While some writers may benefit from AI-driven suggestions, others fear that such tools will undermine interpersonal engagement and diminish the tacit knowledge exchanged through human mentorship. As Fiesler et al.~\cite{fiesler2016archive} suggest, these social dynamics are essential to the cultural sustainability of online communities. Understanding how AI technologies reshape mentorship thus offers a window into the broader reconfiguration of labor, creativity, and collaboration in digital participatory cultures.

Fanfiction communities provide a uniquely rich site for exploring these questions, given their long-standing, self-regulated ecosystems of collective creativity. The responses gathered in this study reflect an ambivalent stance toward AI: some participants embrace its potential to support or inspire, while others express deep concern about its ethical, social, and cultural implications. These tensions highlight the need for a theory of co-creation that accounts not only for technical collaboration between humans and machines but also for the affective, ethical, and communal dimensions of creative practice in digitally mediated environments.

\subsection{Implications for the Design of Online Fanfiction Platforms}
Our findings have implications for the design of online fanfiction platforms. Here, we discuss design interventions to address the concerns raised by fanfiction readers and writers. While our focus is interventions for fanfiction platforms our suggestions could also be applied to other creative communities, which shares values with fanfiction communities \cite{Jenkins2019ArtHN} such as independent publishing platforms \cite{vadde2017amateur}, collaborative writing forums, digital art communities, video-based fan and storytelling platforms \cite{desouza2019fans}, open-source software development communities \cite{winter2021communities}, and music creation platforms.

\subsubsection{Disclosure of AI assistance} Participants expressed significant concerns about the transparency of AI-generated content and the potential for readers to unknowingly engage with AI-authored stories. A core design intervention should involve fine-grained and nuanced labeling of AI-assisted works. Such labeling should reflect the varying degrees of AI involvement and assistance in the writing process. Our results indicate that participants accept the use of AI-powered tools for a variety of writing-related tasks. Several social media platforms, as well as AI-powered writing assistants, have already introduced features for detecting and labeling AI-generated text (e.g. \cite{meta2024ai, vimeo2024ai, grammarly2024ai}). However, we envision a user-driven self-disclosure system, rather than automatic detection, which encourages (and/or requires) users to share both rich context and the specific tools and ways in which they used AI to assist their writing. For example, a non-native English speaker might share that they used GenAI tools to revise their text for clarity and overcome a language barrier. Designing such a system and introducing practices for fine-grained self-disclosure would both satisfy readers' desire for transparency and support writers in highlighting their creative contributions, even when assisted by AI. Such a system could be complemented by AI-detection tools or community-driven labeling mechanisms to protect platforms from bad actors and maintain high-quality, relevant content.

Fanfiction platforms could further utilize such transparency measures by integrating AI usage insights into recommendation systems. For example, labeling could help platforms tailor recommendations based on user preferences for fully human, hybrid, or AI-assisted works, ensuring users feel both informed and engaged. Recommendation systems could also highlight AI-generated content in fandoms with low activity, helping users to discover content and reengage with such fandoms. Such systems could also highlight emerging human authors or non-native English writers who use AI assistance to develop and improve their writing, showcasing their originality and creativity. Thereby, encouraging the preservation of core community values such as creativity and appreciation for human labor.

\subsubsection{Consent for AI Training}
Considering the significant concerns expressed by participants for the integrity of intellectual property, fanfiction platforms should introduce design interventions that allow users to opt in or opt out of using their content in AI training datasets. This feature should be easily accessible and visible. In the absence of such control, many writers have chosen to make their account private~\cite{heather2023ao3}. However, to maintain thriving participatory fanfiction communities, it is important that writers continue to share publicly and engage with community members, trusting the platform that their data will not be used to train AI without their consent.

\subsubsection{Enriching Engagement}
Design interventions for fanfiction platforms could address participants’ concerns regarding potential loss of user engagement by creating opportunities to enrich engagement:

Enlivening Dormant Fandoms: AI tools can breathe new life into inactive or low-activity fandoms. Participants noted that AI could generate content in fandoms with little recent activity, keeping these spaces vibrant and fostering renewed community interest. Platforms could identify such fandoms and promote AI-assisted or encourage human-generated works to rekindle engagement.

Enhancing Content Recommendations: AI-based personalized recommendations that respect user preferences (as discussed above) can bridge gaps between readers and writers, facilitating connections across shared interests.

Valuing Human Labor: A critical design intervention involves highlighting and celebrating human-created works. While some platforms currently implement basic recognition features like user recommendations or story likes, there are additional opportunities for further elevating the role of human creativity. For example, platforms could introduce dynamic achievement badges that evolve with milestones such as ``Best Plot Twists'' or ``Most Improved Writing''. Additionally, platforms could allow readers and writers to nominate works for unique badges like ``Most Emotional Ending'' or ``Funniest One-Liner'' fostering deeper engagement and a sense of collective celebration. Such features would highlight human effort and potentially strengthen the connection.

Increasing Engagement Through Collaboration: AI-powered collaboration tools could enhance mentoring relationships in fanfiction communities, aligning with the ``distributed mentoring'' concept described by Aragon and Davis \cite{aragon2019writers}. For example, GenAI tools could recommend potential collaborators based on shared interest or writing style, facilitating connections between writers who might otherwise not interact. Similarly to \cite{ShaerBrainwriting24}, tools can incorporate shared brainwriting spaces and collaborative prompts to help writers jointly explore creative ideas, thereby fostering a sense of community-driven creativity. Platforms could also acknowledge readers and reviewers who provide critical feedback, offering recognition for their contributions through badges or spotlights. Such features would encourage participation and deepen the bonds within fanfiction communities, ensuring that collaboration remains a core component of engagement.

\subsubsection{Skill Development}
Participants expressed concerns about the potential erosion of writing skills due to over-reliance on AI tools. 
A fanfiction platform could use GenAI to provide further mentoring. Complementing asynchronous distributed mentoring with real-time guidance by a GenAI-powered personal tutor. Such a tutor could guide writers through their writing process, offering advice and applying a Socratic method to help writers develop various skills such as research, planning, characters, and narrative development. Khan Academy has demonstrated the efficacy of such an approach with their Khanmigo tutor, which provides personalized writing coaching \cite{khan2024brave}.

\subsubsection{Fostering Human Creativity}
Finally, design solutions could involve mechanisms to detect and mitigate homogeneity in AI-assisted outputs, ensuring that AI serves as a tool to amplify rather than standardize human creativity. For example, platforms could introduce diverse writing prompts that challenge writers to explore unconventional narratives, genres, or character perspectives. 

Additionally, tools could be designed to prioritize originality over replication, such as by providing feedback that suggests innovative alternatives. Real-time creative metrics could also be integrated, offering insights into the uniqueness of a user's work compared to platform-wide trends~\cite{mokryn2021domain}, thus inspiring users to strive for greater individuality. Through these strategies, platforms can reinforce the value of human creativity while leveraging AI to inspire rather than constrain.

\subsection{Limitations and Future Work}
While our sample of 157 participants offers valuable insights into active fanfiction community members, we acknowledge that it is skewed toward North American and English-speaking participants. This reflects the dominant presence of English-language platforms such as AO3 and the distribution channels available to us, but it limits the generalizability of our findings to global fanfiction communities, especially those operating primarily in non-English contexts. Importantly, this limitation is not only statistical. Cultural norms, access to digital platforms, and local attitudes toward technology can significantly shape how AI is perceived and used in creative communities. For example, fanfiction writers in East Asia~\cite{subin2024fanfiction,prakash2024empire} or Latin America may have distinct expectations around collaboration, attribution, or the appropriateness of integrating AI into the writing process. To fully capture the diversity of perspectives on AI in fanfiction, future research should include more regionally and linguistically diverse participants, allowing for a more nuanced understanding of how cultural context influences the perceived risks and benefits of generative technologies. 

This study has limitations that outline the scope of our findings and indicate potential directions for future research. Our research examines the social dynamics of a creative community during its transition phase, focusing on the evolving conditions, the process of GenAI adoption, and the various concerns arising from these changes. These concerns relate to the community's social fabric, core values, potential loss of control, engagement, and other key aspects. 

While it is important to understand how a dynamic, collaborative, creative community such as fanfiction responds to the growing presence of automation and augmentation tools, capturing the full scope of this transformation requires a longitudinal perspective. A study conducted over time would allow us to observe how creative practices, social norms, and ethical considerations evolve as AI tools become more embedded in the writing process. This includes tracing shifts in attitudes toward authorship, originality, and attribution, as well as concerns around consent, transparency, and the legitimacy of AI-assisted contributions. Importantly, a longitudinal approach would also enable us to examine the roles AI may play both overtly—as visible writing assistants or co-authors—and covertly, through less acknowledged or uncredited use. By combining temporal analysis with in-depth interviews, this future work will help surface subtle changes in peer interactions, feedback mechanisms, and the community’s evolving expectations. Ultimately, such an approach promises a deeper understanding of how generative AI shapes not only creative output but also the ethical, social, and cultural values of participatory storytelling communities.

\section{Conclusions}
We examined fanfiction communities' evolving dynamics as they encounter GenAI integration. Our findings highlight important considerations for how these changes impact community interactions, participatory culture, and the design of online platforms.

First, the presence of AI-generated content is influencing perceptions and interactions within fanfiction communities. While some members see opportunities, such as revitalizing less active fandoms, there are also concerns about how AI might affect values like creativity, authenticity, and the sense of connection among members. Reactions to GenAI range from cautious interest to strong reservations, suggesting that the integration of AI may shift community dynamics in various ways.

Our research then points to the potential challenges AI poses to the participatory nature of fanfiction. As AI tools become more embedded in the writing process, there are concerns about maintaining the collaborative and human-centered aspects of these communities. This underscores the importance of integrating AI in ways that respect and preserve the values of creativity, originality, and community engagement that are central to fanfiction.

Finally, our study offers design implications for online fanfiction platforms, emphasizing the need to address ethical considerations and maintain transparency in the use of AI. Participants expressed concerns about the authenticity of AI-generated content and the need to protect intellectual property. We suggest that platforms could maintain clear disclosures about AI involvement in content creation and provide users with choices about using their content in AI training datasets.

Overall, our findings call for a careful approach to the integration of AI in creative communities, one that balances technological innovation with the preservation of the core values and connections that define these spaces.



\section*{Declarations}
\textbf{Funding:} The authors received no financial support for the research, authorship, or publication of this article. \\ 
\textbf{Conflict of Interest:} The authors declare no conflicts of interest.\\  
\textbf{Ethical Approval:} This research was conducted in compliance with ethical guidelines and received approval from the ethics committees of the University of Haifa (\#47116) and Wellesley College (\#24237R-E).  \\
\textbf{Data Availability:} The data that support the findings of this study are available from the corresponding author, [OM], upon reasonable request.

\bibliographystyle{ACM-Reference-Format}

\newpage
\appendix
\clearpage
\textbf{Appendix A}
\begin{table*}[htbp]
\small
\centering
\caption{Fanfiction survey questions --  demographics and background}
\begin{tabular}{|p{0.07\textwidth}|p{0.43\textwidth}|p{0.35\textwidth}|p{0.1\textwidth}|}
\hline
\textbf{Key} & \textbf{Question} & \textbf{Answers} & \textbf{Participants} \\
\hline
Q2 & How old are you? & \{Under 18\}; \{18-24\}; \{25-34\}; \{35-44\}; \{45-54\}; \{55-64\}; \{65+ years old\} & ALL \\
\hline
Q3 & How do you describe yourself? & \{Female\}; \{Male\}; \{Non-binary / third gender\}; \{Prefer to self-describe\}; \{Prefer not to say\} & ALL \\
\hline
Q4 & What is your level of education? & 
\begin{enumerate}[nosep,leftmargin=*]
\item Undergraduate degree
\item Postgraduate degree
\item Secondary education
\item Vocational or technical training
\item Prefer not to say
\end{enumerate} & ALL \\
\hline
country & List of Countries & All the countries & ALL \\
\hline
Q6 & Primary Language - Selected Choice & \{English\}; \{Spanish\}; \{Japanese\}; \{Other\} & ALL \\
\hline
Q7 & Duration of Involvement in Fanfiction & \{Less than 1 year\}; \{1-5 years\}; \{6-10 years\}; \{More than 10 years\}; \{Prefer not to say\} & ALL \\
\hline
Q8 & On what platforms do you engage with (read or post) fanfiction? & \{Archive Of Our Own\}; \{Wattpad\}; \{Tumblr\}; \{Fanfiction.net\}; \{Discord\}; \{Commaful\}; \{Quotev\}; \{Other\} & ALL \\
\hline
Q9 & On a scale from 1 to 5, where 1 indicates 'Minimal Involvement' and 5 indicates 'Extensive Involvement', to what extent do you engage with fanfiction? & Likert Scale & ALL \\
\hline
Q10 & In regards to your engagement with fanfiction, which of the below options would you say best describes your mode of involvement with the fanfiction community? & 
\begin{enumerate}[nosep,leftmargin=*]
\item Exclusively a reader
\item Currently a reader, hope to write
\item Mostly reader, sometimes writer
\item Equally reader and writer
\item Mostly writer, sometimes reader
\item Currently writer, read in past
\item Exclusively a writer
\item Other
\end{enumerate} & ALL \\
\hline
Q12 & Fandoms of Interest & Open Text Responses & ALL \\
\hline
Q13 & What key elements do you believe contribute to the quality of fanfiction writing? (Please choose up to three.) &
\begin{enumerate}[nosep,leftmargin=*]
\item Authenticity to the source material
\item Originality and creativity in plot and character development
\item Emotional depth and character introspection 
\item  Engaging and coherent narrative structure
\item Well-crafted dialogue
\item Accurate grammar and stylistic consistency, Other 
\end{enumerate} & ALL \\
\hline
\end{tabular}
\label{questP1}
\end{table*}
\newpage

\begin{table}[htbp]
\centering
\caption{Fanfiction survey questions for readers part 1}
\small
\begin{tabular}{|p{0.05\textwidth}|p{0.43\textwidth}|p{0.35\textwidth}|p{0.1\textwidth}|}
\hline
\textbf{Key} & \textbf{Question} & \textbf{Answers} & \textbf{Participants} \\
\hline
Q14 & Do you read fanfiction? & \{Yes\}; \{No\} & Readers \\
\hline
Q15 & What first drew you into the fanfiction community? & 
\begin{enumerate}[nosep,leftmargin=*]
\item Wanting to spend more time with favorite characters
\item Wanting to spend more time with the world/worldbuilding
\item Interest in exploring non-canonical relationships
\item Interest in exploring canonical relationships
\item The diversity of storytelling
\item Ease of access to fanfiction
\item Wanting to connect with others
\item Other
\end{enumerate} & Readers \\
\hline
Q16 & How do you navigate finding and choosing fanfiction to read? & 
\begin{enumerate}[nosep,leftmargin=*]
\item Most popular fics
\item Recently-published/updated fic
\item Familiar fandoms or characters
\item Canon-blind reading
\item Specific tropes, plots, or AU settings
\item Specific relationships and pairings
\item Specific characters
\item Recommended by others
\item Known authors
\item New authors
\item Other
\end{enumerate} & Readers \\
\hline
Q17 & What factors influence your decision to select one fanfiction story over another for reading? & Open Text Responses & Readers \\
\hline
Q18 & How often do you leave a kudos/like? & \{Almost Never\}; \{Not Often\}; \{Sometimes\}; \{Often\}; \{Very Often\} & Readers \\
\hline
Q20 & How often do you leave a comment? & \{Almost Never\}; \{Not Often\}; \{Sometimes\}; \{Often\}; \{Very Often\} & Readers \\
\hline
Q19 & Have you encountered fanfiction stories that you knew were generated by AI? & \{Yes\}; \{Maybe\}; \{No\} & Readers \\
\hline
Q21 & How would your perception of a story change if you discovered it was AI-written only after you had finished reading it? & \{Unchanged\}; \{More Positive\}; \{More Negative\}; \{Deceived\}; \{Surprised\}; \{Other\} & Readers \\
\hline
Q22 & How would knowing in advance that a fanfiction story was generated by AI influence your interest in reading it? & \{Increase interest\}; \{No influence\}; \{Decrease interest\}; \{Depends on the context\}; \{Other\} & Readers \\
\hline
Q23 & When engaging with fanfiction, how often do you think about the possibility that the fic was written with the help of GenAI? & \{Never\}; \{Sometimes\}; \{About half the time\}; \{Most of the time\}; \{Always\} & Readers \\
\hline
Q24 & Has the increased use of GenAI impacted how often you leave a like/kudos or comment? & \{Increased Engagement\}; \{No change\}; \{Decreased Engagement\}; \{Depends on the content quality\}; \{Other\} & Readers \\
\hline
\end{tabular}
\label{questP2}
\end{table}
\clearpage

\begin{table}[htbp]
\centering
\caption{Fanfiction survey questions for readers part 2}
\small
\begin{tabular}{|p{0.06\textwidth}|p{0.4\textwidth}|p{0.32\textwidth}|p{0.1\textwidth}|}
\hline
\textbf{Key} & \textbf{Question} & \textbf{Answers} & \textbf{Participants} \\
\hline
Q25 & When reading AI-generated fanfiction, what aspects do you pay most attention to in assessing its quality? (Select all that apply) &
\begin{enumerate}[nosep,leftmargin=*]
\item Character development
\item Plot coherence
\item Emotional depth
\item Authenticity to the source material
\item Creativity and originality
\item Ease of access to fanfiction
\item Writing style and fluency
\item Other\}
\end{enumerate}& Readers \\
\hline
Q26 & How important is it to you that characters will stay true to their original form and character? & \{Not at all important\}; \{Slightly important\}; \{Moderately important\}; \{Very important\}; \{Extremely important\} & Readers\\
\hline
Q27\_1 & I would read the fic regardless of how the author had used GenAI & \{Strongly disagree\}; \{Somewhat disagree\}; \{Neither agree nor disagree\}; \{Somewhat agree\}; \{Strongly agree\} & Readers \\
\hline
Q27\_2 & I would be willing to read a fic if I knew the author created the entire story using AI and made little to no edits before sharing it & \{Strongly disagree\}; \{Somewhat disagree\}; \{Neither agree nor disagree\}; \{Somewhat agree\}; \{Strongly agree\} & Readers \\
\hline
Q27\_3 & I would read the fic knowing the author produced a first draft with AI and manually made minor revisions/rewrites & \{Strongly disagree\}; \{Somewhat disagree\}; \{Neither agree nor disagree\}; \{Somewhat agree\}; \{Strongly agree\} & Readers \\
\hline
Q27\_4 & I would consider reading a fic if I knew the author generated the initial draft with AI and then personally conducted major revisions and rewrites & \{Strongly disagree\}; \{Somewhat disagree\}; \{Neither agree nor disagree\}; \{Somewhat agree\}; \{Strongly agree\} & Readers \\
\hline
Q27\_5 & I would be open to reading a fic knowing that the author initially crafted it by hand and then used AI for significant revisions & \{Strongly disagree\}; \{Somewhat disagree\}; \{Neither agree nor disagree\}; \{Somewhat agree\}; \{Strongly agree\} & Readers \\
\hline
Q27\_6 & I would engage with a fic if the author utilized AI tools strictly for grammar and spell checking, but not beyond that scope & \{Strongly disagree\}; \{Somewhat disagree\}; \{Neither agree nor disagree\}; \{Somewhat agree\}; \{Strongly agree\} & Readers \\
\hline
Q27\_7 & I would read the fic knowing the author first drafted manually and used AI to make minor revisions & \{Strongly disagree\}; \{Somewhat disagree\}; \{Neither agree nor disagree\}; \{Somewhat agree\}; \{Strongly agree\} & Readers \\
\hline
Q27\_8 & I would steer clear of any fic where the author has employed GenAI at any stage of the writing process & \{Strongly disagree\}; \{Somewhat disagree\}; \{Neither agree nor disagree\};  \{Somewhat agree\}; \{Strongly agree\} & Readers \\
\hline
Q29 & Has the notion that a fic might have been created using GenAI influenced your interactions within the fandom community? If yes, could you share how? & Open Text Responses & Readers \\
\hline
Q30 & What particular markers or cues lead you to suspect a story might have been composed with the assistance of GenAI? & Open Text Responses & Readers \\
\hline
\end{tabular}
\label{questP3}
\end{table}

\clearpage
\begin{table}[htbp]
\centering
\caption{Fanfiction survey questions for readers part 3}
\small
\begin{tabular}{|p{0.05\textwidth}|p{0.43\textwidth}|p{0.35\textwidth}|p{0.09\textwidth}|}
\hline
\textbf{Key} & \textbf{Question} & \textbf{Answers} & \textbf{Participants} \\
\hline
Q44 & How do you feel about the concept of increased integration of AI into fanfiction writing? &
\{Excited\};
\{Mostly open-minded but cautious\};
\{Indifferent or unsure\};
\{Mostly cautious but open-minded\};
\{Opposed\};
\{Depends on the implementation\};\{Other\}
& ALL\\
\hline
Q45 & To what extent do you agree: AI-generated content can maintain the authenticity of fanfiction narratives while offering innovative storytelling? & \{Strongly disagree\}; \{Somewhat disagree\}; \{Neither agree nor disagree\}; \{Somewhat agree\}; \{Strongly agree\} & ALL \\
\hline
Q46 & What is your biggest hope or concern regarding the intersection of AI and fanfiction? & Open Text Responses & ALL \\
\hline
Q47 & In your view, are there measures that could be taken to ensure that any integration of AI enhances rather than dilutes the fanfiction community's spirit? If so, what? & Open Text Responses & ALL \\
\hline
Q48\_1& To what extent do you agree or disagree with the following statement: AI-generated fanfiction can never replicate the emotional nuance and depth that comes from human authors?& \{Strongly disagree\}; \{Somewhat disagree\}; \{Neither agree nor disagree\}; \{Somewhat agree\}; \{Strongly agree & ALL \\
\hline
Q48\_2& To what extent do you agree or disagree with the following statement: Authors should be transparent about the involvement of AI in their creative writing process to ensure transparency is maintained with their readers?& \{Strongly disagree\}; \{Somewhat disagree\}; \{Neither agree nor disagree\}; \{Somewhat agree\}; \{Strongly agree & ALL \\
\hline
Q48\_3& To what extent do you agree or disagree with the following statement: Reading fanfiction is a social, community-centric activity?& \{Strongly disagree\}; \{Somewhat disagree\}; \{Neither agree nor disagree\}; \{Somewhat agree\}; \{Strongly agree & ALL \\
\hline
Q48\_4& To what extent do you agree or disagree with the following statement: Writing fanfiction is a social, community-centric activity?& \{Strongly disagree\}; \{Somewhat disagree\}; \{Neither agree nor disagree\}; \{Somewhat agree\}; \{Strongly agree & ALL \\
\hline
Q48\_5& To what extent do you agree or disagree with the following statement: The emergence of AI-authored fanfiction poses a danger to the social aspects of the fanfiction community?& \{Strongly disagree\}; \{Somewhat disagree\}; \{Neither agree nor disagree\}; \{Somewhat agree\}; \{Strongly agree & ALL \\
\hline
Q48\_6& To what extent do you agree or disagree with the following statement: An influx of AI-created works might inundate fanfiction platforms, potentially overshadowing human-authored stories?& \{Strongly disagree\}; \{Somewhat disagree\}; \{Neither agree nor disagree\}; \{Somewhat agree\}; \{Strongly agree & ALL \\
\hline
Q48\_7& To what extent do you agree or disagree with the following statement: Fanfiction is a space for human creativity to flourish & \{Strongly disagree\}; \{Somewhat disagree\}; \{Neither agree nor disagree\}; \{Somewhat agree\}; \{Strongly agree & ALL \\
\hline
Q48\_8& To what extent do you agree or disagree with the following statement: Relying on AI for content creation could stifle human creativity and creative freedom?& \{Strongly disagree\}; \{Somewhat disagree\}; \{Neither agree nor disagree\}; \{Somewhat agree\}; \{Strongly agree & ALL \\
\hline
Q48\_9 & To what extent do you agree or disagree with the following statement: Worries surrounding AI's role in fanfiction might be exaggerated?& \{Strongly disagree\}; \{Somewhat disagree\}; \{Neither agree nor disagree\}; \{Somewhat agree\}; \{Strongly agree & ALL \\
\hline
Q48\_10 & To what extent do you agree or disagree with the following statement: The perceived depth or quality of a story by its reader holds greater significance than the method of its creation by its author? & \{Strongly disagree\}; \{Somewhat disagree\}; \{Neither agree nor disagree\}; \{Somewhat agree\}; \{Strongly agree & ALL \\
\hline
Q49 & How accurately do you believe you can identify a story written with GenAI? & \{Not at all accurately\}; \{Not very accurately\}; \{Neutral / Unsure\}; \{Somewhat accurately\}; \{Very accurately\} & ALL \\
\hline
Q51 & What concerns, if any, do you have about using AI in fanfiction writing? (Select all that apply) & [Same options as Q35] & ALL \\
\hline
\end{tabular}
\label{questP4}
\end{table}

\clearpage
\begin{table}[htb]
\centering
\caption{Fanfiction survey questions for writers part 1}
\small
\begin{tabular}{|p{0.04\textwidth}|p{0.24\textwidth}|p{0.48\textwidth}|p{0.1\textwidth}|}
\hline
\textbf{Key} & \textbf{Question Text} & \textbf{Options of Answers} & \textbf{Participants} \\
\hline
Q31 & Do you write fanfiction? & Yes, No & Writers \\
\hline
Q32 & What drives you to write fanfiction? (Select all that apply) & 
\begin{enumerate}[nosep,leftmargin=*]
\item Exploring favorite characters
\item Exploring world/worldbuilding
\item Exploring non-canonical relationships
\item Exploring canonical relationships
\item Freedom to write preferred topics
\item Ease of entry into community
\item Connecting with other fans
\item Developing writing skills
\item Other
\end{enumerate} & Writers\\
\hline
Q33 & How do you currently integrate technology or digital tools into your writing process? (Select all that apply) & 
\begin{enumerate}[nosep,leftmargin=*]
\item Writing software
\item Online research
\item Digital note-taking apps
\item Traditional methods\
\item AI-powered writing assistants
\item Other
\end{enumerate} & Writers \\
\hline
Q34 & In your own words, please describe your writing process: & Open Text Responses & Writers \\
\hline
Q35 & What concerns, if any, do you have about using AI in fanfiction writing? (Select all that apply) & 
\begin{enumerate}[nosep,leftmargin=*]
\item Loss of personal touch
\item Over-reliance on technology
\item Copyright infringement
\item Inaccurate character portrayal
\item Ethical implications
\item Low-quality content
\item Impact on emerging writers
\item Loss of personal journey
\item Other
\end{enumerate} & Writers \\
\hline
Q36 & In your own words, tell us about any concerns that you might have: & Open Text Responses & Writers \\
\hline
Q38 & If you do use AI-powered tools in your writing - do you acknowledge its use in your published work in some way? Please explain your thoughts on this topic: & Open Text Responses & Writers \\
\hline
Q39  & In your writing process, how do you view the role of AI? & 
\begin{enumerate}[nosep,leftmargin=*]
\item As a co-creator involved in content generation
\item As a tool or assistant for specific tasks
\item As both co-creator and tool/assistant
\item As neither co-creator nor tool/assistant
\item A tool that others use that I have to compete with
\item Other
\end{enumerate} & Writers \\
\hline
\end{tabular}
\label{questP5}
\end{table}

\clearpage
\begin{table}[htb]
\centering
\caption{Fanfiction survey questions for writers part 2}
\small
\begin{tabular}{|p{0.04\textwidth}|p{0.34\textwidth}|p{0.38\textwidth}|p{0.1\textwidth}|}
\hline
\textbf{Key} & \textbf{Question Text} & \textbf{Options of Answers} & \textbf{Participants} \\
\hline
Q40\_1 - Q40\_11 & Familiarity with writing tools (ChatGPT, Grammarly, Jasper, QuillBot, Wordtune, Claude, NovelCrafter, AutoCrit, SudoWrite, Gemini, Other) & \{Never heard of it\}; \{Know it but haven't used it\}; \{Used it a lot\}; \{Used it a few times\} & Writers \\
\hline
Q41\_1 & To what extent do you agree or disagree with the following statements: - It's ok to first draft the writing manually then use AI for minor revisions & \{Strongly disagree\}; \{Somewhat disagree\}; \{Neither agree nor disagree\}; \{Somewhat agree\}; \{Strongly agree\} & Writers\\
\hline
Q41\_2 & To what extent do you agree or disagree with the following statements: - It's ok to first draft the writing manually then use AI for major revisions & \{Strongly disagree\}; \{Somewhat disagree\}; \{Neither agree nor disagree\}; \{Somewhat agree\}; \{Strongly agree\} & Writers\\
\hline
Q41\_3 & To what extent do you agree or disagree with the following statements: - It's ok to produce a first draft with AI and manually make minor revisions/rewrites & \{Strongly disagree\}; \{Somewhat disagree\}; \{Neither agree nor disagree\}; \{Somewhat agree\}; \{Strongly agree\} & Writers\\
\hline
Q41\_4 & To what extent do you agree or disagree with the following statements: - It's ok to produce a first draft with AI and manually make major revisions/rewrites & \{Strongly disagree\}; \{Somewhat disagree\}; \{Neither agree nor disagree\}; \{Somewhat agree\}; \{Strongly agree\} & Writers\\
\hline
Q41\_5 & To what extent do you agree or disagree with the following statements: - It's ok to produce the entire fic with AI and make minimal to no edits before publishing& \{Strongly disagree\}; \{Somewhat disagree\}; \{Neither agree nor disagree\}; \{Somewhat agree\}; \{Strongly agree\} & Writers\\
\hline
Q42 & In which writing stages is it ok to collaborate with AI? (Select all that apply) &
\{Coming Up with Idea\}; \{Plotting, Writing\}; \{Editing\}; \{None of the above\}; \{Other\}& Writers \\
\hline
Q43 & In which aspects of your writing do you envision AI would be most beneficial? &
\begin{enumerate}[nosep,leftmargin=*]
\item Generating ideas and inspirations
\item Plot and character development assistance
\item Real-time feedback and suggestions
\item Overcoming writer's block
\item Drafting and outlining
\item Editing and proofreading
\item Character and world-building
\item Don't use AI at all
\item Other
\end{enumerate}& Writers \\
\hline
\end{tabular}
\label{questP6}
\end{table}
\end{document}